\documentclass[12pt, draftclsnofoot, onecolumn]{IEEEtran}

\usepackage{graphicx}

\usepackage{subfigure}
\usepackage{cite}
\usepackage{amsmath,amssymb,amsfonts,bbm}
\DeclareMathOperator*{\argmax}{argmax} 
\usepackage{textcomp}
\usepackage{xcolor}
\usepackage{comment}
\def\BibTeX{{\rm B\kern-.05em{\sc i\kern-.025em b}\kern-.08em
    T\kern-.1667em\lower.7ex\hbox{E}\kern-.125emX}}
\newcommand\mydots{\hbox to 0.75em{.\hss.\hss.}}

\allowdisplaybreaks
\usepackage{stfloats}
\usepackage{cuted}
\usepackage[detect-all=true]{siunitx}
\DeclareSIUnit\decibelm{dBm}
\DeclareSIUnit\packets{packets}
\DeclareSIUnit\pct{percentile}
\AtBeginDocument{
	\DeclareSIUnit\bit{b}
	\DeclareSIUnit\bitpersec{bps}

	\newcolumntype{C}[1]{>{\centering\let\newline\\\arraybackslash\hspace{0pt}}m{#1}}
	\newcolumntype{L}[1]{>{\raggedright\let\newline\\\arraybackslash\hspace{0pt}}m{#1}}
	
}
\DeclareSIUnit[per-mode=symbol, per-symbol=p]\bps{\bit \per \second}
\DeclareSIUnit[per-mode=symbol, per-symbol=p]\kbps{\kilo \bit \per \second}\DeclareSIUnit[per-mode=symbol, per-symbol=p]\Mbps{\mega \bit \per \second}
\DeclareSIUnit[per-mode=symbol]\bpsHz{\bitpersec \per \hertz}
\usepackage{soul}

\usepackage{booktabs}
\usepackage{amsthm}
\usepackage{ textcomp }
\usepackage{subfigure}
\usepackage{bm}
\usepackage{multirow}
\usepackage{longtable}
\usepackage{amssymb}
\usepackage{xcolor}
\usepackage{algorithm,algorithmic}
\usepackage[algo2e,ruled,vlined]{algorithm2e}

\ifCLASSOPTIONcompsoc
  \usepackage[nocompress]{cite}
\else
  \usepackage{cite}
\fi
\usepackage{comment}

\begin{document}
\bstctlcite{IEEEexample:BSTcontrol}
\graphicspath{{./Figures/}}

\title{Extended Reality over 3GPP 5G-Advanced New Radio: Link Adaptation Enhancements}

\author{
{Pouria Paymard}, 
{Abolfazl Amiri}, 
{Troels E. Kolding}, and
{Klaus I. Pedersen}

\thanks{P. Paymard and K.I. Pedersen are with the Department of Electronic Systems,  Technical Faculty of IT and Design; Aalborg University,	Denmark; E-mail: { pouriap@es.aau.dk}}

\thanks{A.Amiri, T. E. Kolding and K. I. Pedersen are with Nokia, Aalborg,	Denmark; E-mail: { abolfazl.amiri@nokia.com,\{klaus.pedersen, troels.kolding\}@nokia-bell-labs.com}}
}

\maketitle


\begin{abstract}
One of the rapidly emerging services for  fifth-generation (5G)-Advanced is eXtended Reality (XR) which combines several immersive experiences and cloud gaming services. Those services are demanding as they call for relatively high data rates under tight latency constraints, sometimes also referred to as dependable real-time applications. Supporting as many XR users per cell requires highly efficient radio solutions. In this paper, we propose an enhanced channel quality indicator (CQI) that results in a better link adaptation to unleash the full performance potential of code block group (CBG) based transmissions for XR cases. We present both an analytical analysis of the related problems and solutions, as well as an extensive dynamic system-level performance assessment in line with the 3\textsuperscript{rd} generation partnership project (3GPP)-defined advanced simulation methodologies. Our results show an increased XR system capacity of 17\% to 33\% as compared to what can be supported by current 5G systems with baseline CQI schemes. We also present enhanced CQI complexity-reducing techniques based on derived closed-form expressions that are attractive to the user equipment (UE) implementation.
\end{abstract}

\section{Introduction}\label{sec:introduction}

    Standardization of the fifth generation (5G) cellular, known as 5G New Radio (NR), by the 3rd generation partnership project (3GPP) was first realized by Release-15. Currently, 3GPP is working on Release-18 for 5G-Advanced, which introduces several key enhancements and support for services, adopting research findings from both industry and academia. One of the hot topics considered is extended reality (XR) \cite{3gpp.38.838, petrov2022extended, RP-213587, chen2021frame, MAC_VR}. XR is an umbrella term for three popular immersive applications i.e., virtual reality (VR), augmented reality (AR), and mixed reality (MR) \cite{3gpp.38.838}. 5G-Advanced aims an offering superior XR experience in public, consumer, and industrial sectors.  XR services are characterized by strict quality of service (QoS) requirements. As an example, typical XR services call for data rates from 30 Mbps to 60 Mbps with latency constraints from 5 ms to 15 ms for the radio access network part and reliability targets on the order of 99\% to 99.999\% \cite{3gpp.38.838}. 
    
    In fulfilling these requirements for as many XR users as possible per cell, packet scheduling and link adaptation (LA) play a crucial role in assigning the right amount of radio resources and modulation schemes to users to fulfill the XR service requirements. LA assigns the modulation and the coding rate of the error correction scheme based on the quality of the radio link. Two fundamental components of the LA are the channel quality indicator (CQI) and outer loop LA (OLLA) which work hand in hand to assess the quality of the radio link and assign proper resources to satisfy a certain utility function
    \cite{Guillermo_2020_Channel_Quality_Feedback, ML_Approach_CQI_Feedback, sarret2015dynamicOLLA, blanquez2016eolla, OLLA2015self}.
    
    5G NR supports code block group (CBG)-based transmissions and associated hybrid automatic repeat request (HARQ) operation which divides a transport block (TB) into smaller groups of code blocks (CBs) to maximize the radio resource usage efficiency in HARQ retransmissions.
    Basically, each CB has a cyclic redundancy check (CRC). 
    If one or more CBs in a CBG get failed, the erroneous CBG should be retransmitted as HARQ acknowledgment (ACK) or negative-acknowledgment (NACK) bits are provided per CBG in a TB.
    CBG-based transmissions are essentially aligned with the transmission of huge payloads such as XR cases. This can effectively reduce the retransmission payload size and, accordingly, improve the resource efficiency \cite{khosravirad2017flexible, yeo2017partial, Reduced_CBG_HARQ_for_coexistence_with_urllc_traffic, wu2018multilevel, GlobeCom_eOLLA}.

\subsection{State of the Art}
    
    In the recent 3GPP Release-17 study item on XR over NR \cite{3gpp.38.838}, the basic modeling and system-level evaluation methodologies for XR were agreed, including the definition of several XR-specific performance indicators (KPI). This study item concluded that the current 5G network can support XR services, but also identified several directions of possible enhancements to further boot the XR capacity. A general overview of 3GPP XR research is also available in\cite{petrov2022extended}. In 3GPP 5G-Advanced Release 18, 3GPP is now pursuing the introduction of further XR enhancements as outlined in \cite{RP-213587}. Among others, it includes capacity where enhanced LA and enhanced CQI (eCQI) are under study. 

The open literature is rich in LA studies, hence it would be too exhaustive to list here. In the following, we therefore only summarize a representative subset of those that are most relevant for this article. As an example, in \cite{Guillermo_2020_Channel_Quality_Feedback} the authors address CQI enhancements that entail biased interference filtering of the collected channel quality measurements and so-called Worst-M CQI reporting formats.  The study in \cite{ML_Approach_CQI_Feedback} proposes a machine learning-based approach to address the CQI feedback delay problem.
However, as will be discussed in greater detail in this paper, currently known CQI designs are not designed to take full advantage of CBG-based transmissions for XR use cases. OLLA schemes that operate with traditional HARQ with a single bit ACK/NACK per transport block are studied in \cite{sarret2015dynamicOLLA, blanquez2016eolla, OLLA2015self}.  The work in \cite{sarret2015dynamicOLLA} introduces an algorithm to improve the link robustness to the signal-to-interference-plus-noise ratio (SINR) variability for a 5G centimeter-wave concept.  Other OLLA approaches are studied in \cite{blanquez2016eolla}. The solution in \cite{OLLA2015self} proposes a self-optimization algorithm to adjust the OLLA initial offset. 

There are also several studies of CBG-based transmissions in the literature. As an example, the authors in \cite{khosravirad2017flexible} , and \cite{yeo2017partial} propose CBG-based multi-bit feedback approaches. The study in \cite{Reduced_CBG_HARQ_for_coexistence_with_urllc_traffic} presents new schemes which provide improvements to the current CB/CBG HARQ feedback scheme.  In \cite{wu2018multilevel}, the authors introduced a novel multi-level CBG-based HARQ scheme, which also offers reduced overhead of HARQ feedback.  In \cite{GlobeCom_eOLLA}, we proposed two new low-complexity OLLA algorithms tailored to CBG-based transmissions with multi-bit HARQ feedback for the XR downlink.  Contrary to traditional OLLA algorithms for controlling the first transport block transmission error rate of 10\%,  we aimed at controlling the desired  CBG error rate for the first and the second transmissions. Adopting such solutions was shown to help enhance the XR capacity.
  
\subsection{Contributions}
    In this paper, we propose an eCQI scheme for XR use cases with  CBG-based transmissions. Our main contributions are:
     \begin{itemize}
        \item We present an analytical analysis of the CBG-based transmissions to motivate its potential advantages for the XR use cases and to determine the desired block error rate operating point. This includes cases where errors on individual CBGs are not always independent and identically distributed (i.i.d.), but also cases with a correlation of different error levels.
        \item We present a new eCQI scheme that guides the Next Generation NodeB (gNB) for which modulation and coding scheme (MCS) to use, subject to the actual CBG-based performance at the terminal.
        \item Closed-form expressions for calculating eCQI in the terminal are derived that are attractive for a low complexity implementation. 
        \item We assess the XR performance  under realistic system-level conditions with multiple users, multiple cells, dynamic XR traffic, and accurate modeling of the major performance-determining radio access network functionalities in line with \cite{3gpp.38.838}. The presented results confirm that there are promising benefits to gain in terms of higher XR system capacity by adopting CBG-based HARQ schemes with the proposed eCQI scheme.
     \end{itemize}

\subsection{Structure of the Paper}
    The rest of this paper is organized as follows. 
    Section \ref{P2:Sec2:System Model} describes the network model and XR traffic model. 
    Section \ref{Analytical_Assessment} states an analytical assessment of the CBG-based HARQ retransmissions.
    In Section \ref{CQI enhancements for CBG}, we propose the CBG-based.
    Section \ref{Computational_Complexity_Reduction} discusses the computational complexity of enhanced CQI reporting and closed-form expressions to reduce the added complexity.
    Section \ref{Simulation_Results} includes the simulation setup and the performance evaluation.
    Finally, Section \ref{Conclusion} concludes the paper.

\section{Setting The Scene}\label{P2:Sec2:System Model}

\subsection{Deployment Scenario and Frame Structure}
	We adopt the agreed 3GPP XR system model and related evaluation methodologies in \cite{3gpp.38.838}, assuming a time division duplex (TDD) 5G network with orthogonal frequency-division multiple access (OFDMA). The considered scenario is the indoor hotspot (InH) deployment with $B$ gNBs. We focus on the downlink (DL) performance where we assume 100 MHz carrier bandwidth at 4 GHz, 30 kHz sub-carrier spacing (SCS), and physical resource blocks (PRBs) consisting of 12 sub-carriers \cite{3gpp.38.211}. Each slot consists of 14 orthogonal frequency-division multiplexing (OFDM) symbols. A fixed TDD radio frame configuration with a DDDSU slot pattern with a 2.5 ms repetition period is assumed, where D, S, and U denote Downlink-, Uplink-, and Special-slots. 

\subsection{XR-Specific 3GPP Traffic Model}
	\begin{figure}[t]
		\centerline{\includegraphics[width=0.5\linewidth,trim={3cm 0cm 0cm 0cm},clip]{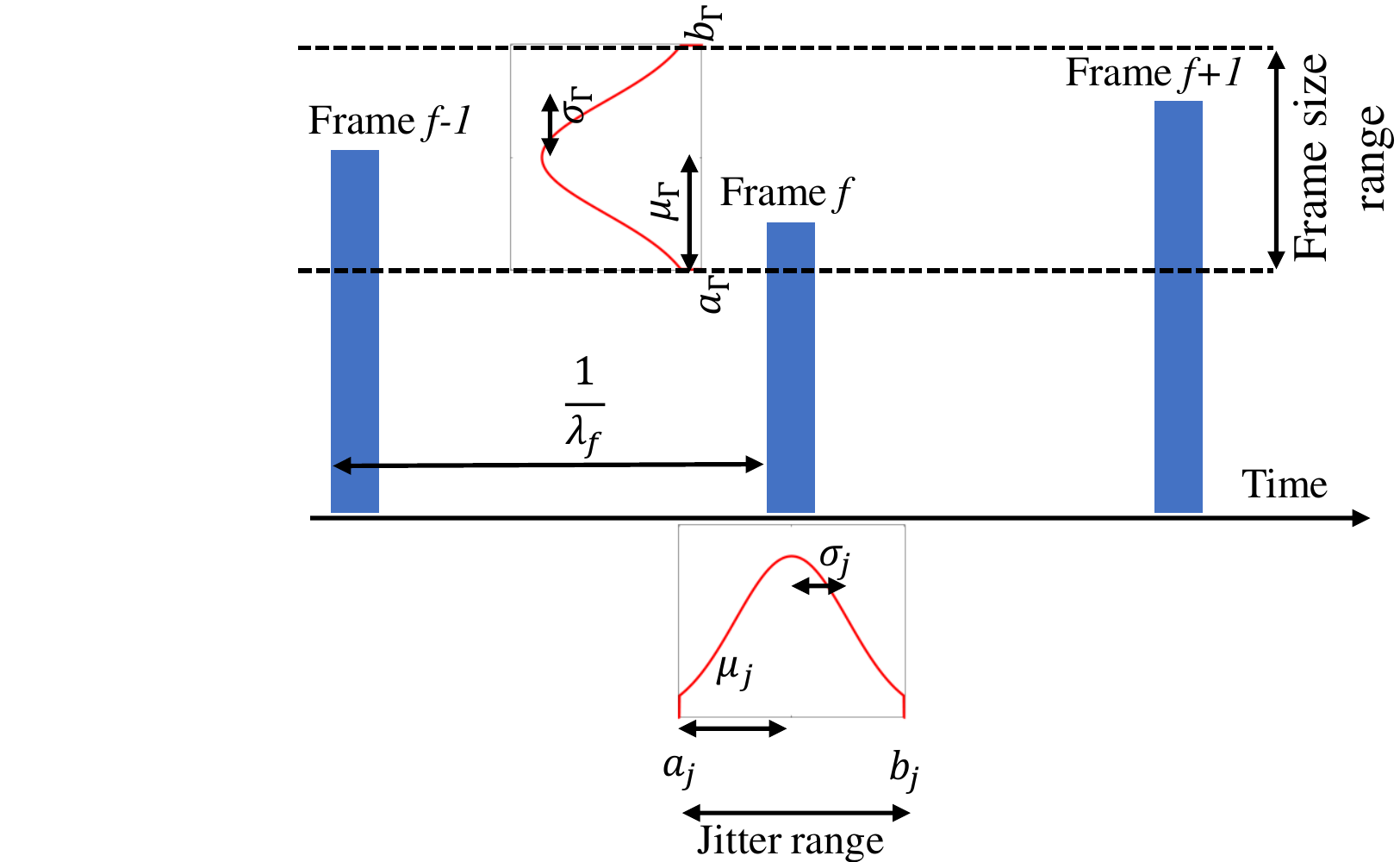}}
		\caption{The downlink XR traffic model according to Release 17 agreements in \cite{3gpp.38.838}.}
		\label{P2:Fig:XR_Traffic_Model}
	\end{figure}
	The assumed downlink XR traffic modeling mimics a case where the XR application server for view-port rendering or other session techniques is performed. Then, the traffic is delivered to XR user equipment (UEs) via 5G the radio access network part in close proximity to the users. XR video frames are periodically generated at the application server based on a fixed  rate of $\lambda_f$ frames per second (fps). The arrival time of XR video frames to gNB is quasi-periodic because of a random jitter that originates from encoding, compressing, routing, etc. The random jitter is denoted by $J_f$, which follows a truncated Gaussian distribution $J_f\sim \mathcal{TN}(\mu_j,\sigma_j,a_j,b_j)$ with mean $\mu_j$, variance $\sigma_j^{2}$, and non-zero interval $[a_j,b_j], a_j \leq b_j$. 	The arrival time of video frame $f$ is expressed as
	\begin{equation}\label{}
	    T_f= f \times \frac{1}{\lambda_f} \times 1000 + J_f.
	\end{equation}
The XR video frames have variable sizes due to the applied video compression algorithms. The video frame size is modeled by a truncated Gaussian distribution $\Gamma_f\sim \mathcal{TN}(\mu_\Gamma,\sigma_\Gamma,a_\Gamma,b_\Gamma)$ with mean $\mu_\Gamma$, variance $\sigma_\Gamma^{2}$, and non-zero interval $[a_\Gamma,b_\Gamma], a_\Gamma \leq b_\Gamma$.
	Throughout the rest of the paper, a video frame simply is denoted as a packet. The assumed XR traffic is illustrated in Fig. \ref{P2:Fig:XR_Traffic_Model}.

\subsection{Radio Resource Scheduling}
Scheduled transmissions are assumed, where each gNB dynamically schedules its users whenever there are pending data in its buffer to be transmitted \cite{PS_5G_Klaus}. If there are pending HARQ retransmissions, those are always prioritized over the transmission of new data. Frequency domain multiplexing of users on PRB resolution, using the well-known proportional-fair (PF) scheduler with the following scheduling metric:
	\begin{equation}\label{P2:eq:PF_metric}
	    u^\ast= \argmax_u \left\{\frac{\psi_{u,\chi}^r}{\Bar{\psi}_u[t]} \right\},
	\end{equation}
where $t$ and $\chi$ are the transmission time interval (TTI) and PRB indices, while  $\Bar{\psi}_u[t]$  is the average delivered throughput to UE $u$ in the past. 	Moreover, $\psi_{u,\chi}^r$ is an estimate of the instantaneous supported data rate of UE $u$ on PRB $\chi$ with the MCS index $r$. The value of $\psi_{u,\chi}^r$ is obtained from the CQI sent by each UE. Notice here that the baseline CQI expresses  the highest data rate $\psi_{u,\chi}^r$ that can be supported with a TBER less than $P_\text{Target}=10\%$. Finally, Chase Combining HARQ re-transmissions \cite{Holma_5G} are assumed as the baseline.
    
\begin{figure}[t]
		\centerline{\includegraphics[width=0.65\linewidth,trim={0cm 0cm 0cm 0cm},clip]{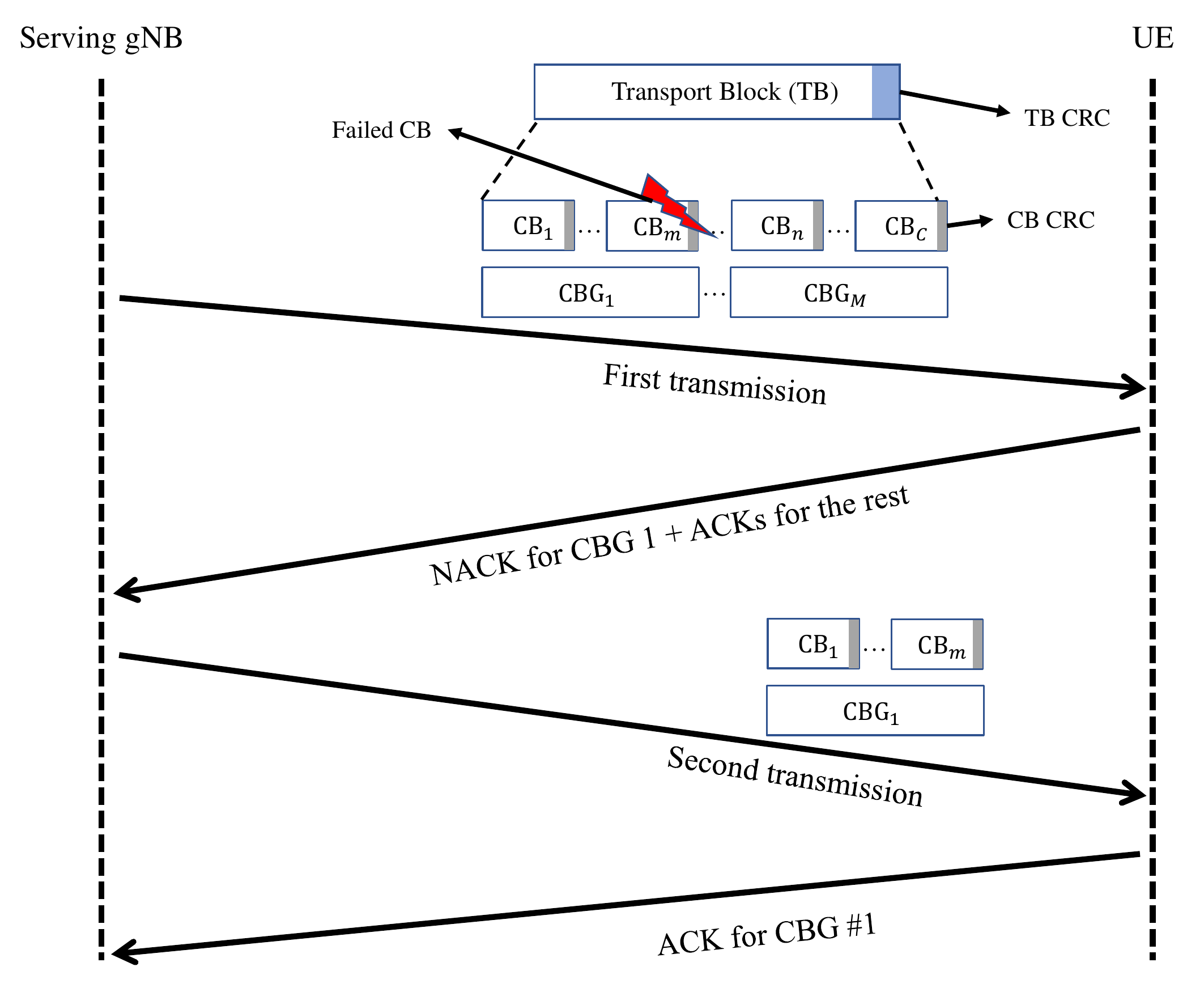}}
		\caption{The construction of $M$ CBGs per TB, each consisting of $C$ CBs.}
		\label{P1:Fig:TB_to_CBG}
	\end{figure}

For the cases with CBG-based transmissions, each scheduled transport block (TB) is divided into multiple CBs that are separately encoded and have their own CRC \cite{3gpp.38.214}. The maximum size of a CB is 8448 bits. These CBs are grouped into CBGs. For each received TB, the receiver provides a multi-bit HARQ feedback that indicates which CBGs are in error, and only the erroneously received CBGs are thereafter retransmitted. A CBG is in error if one, or more, of the CBs are erroneously decoded.  The relationship between TB, CB, CBG, and associated HARQ feedback bits for each CBG is illustrated in Fig. \ref{P1:Fig:TB_to_CBG}.
As defined in the 3GPP technical specification (TS) \cite{3gpp.38.214}, if a UE is configured to receive CBG-based transmissions, the UE obtains the number of CBGs for a TB as $M = \min(F,C)$, where $F$ the maximum number of CBGs per TB is configurable by radio resource control (RRC) as $F \in \left\lbrace 2, 4, 6, 8 \right\rbrace$ for the physical downlink shared channel (PDSCH), and $C$ is the number of CBs in the TB. For XR use cases with the described traffic model, we typically have larger TB sizes with $8$ CBGs per scheduled transmission.

\subsection{Link Adaptation Mechanism}
\begin{figure}[t]
    		\centerline{\includegraphics[width=0.65\linewidth]{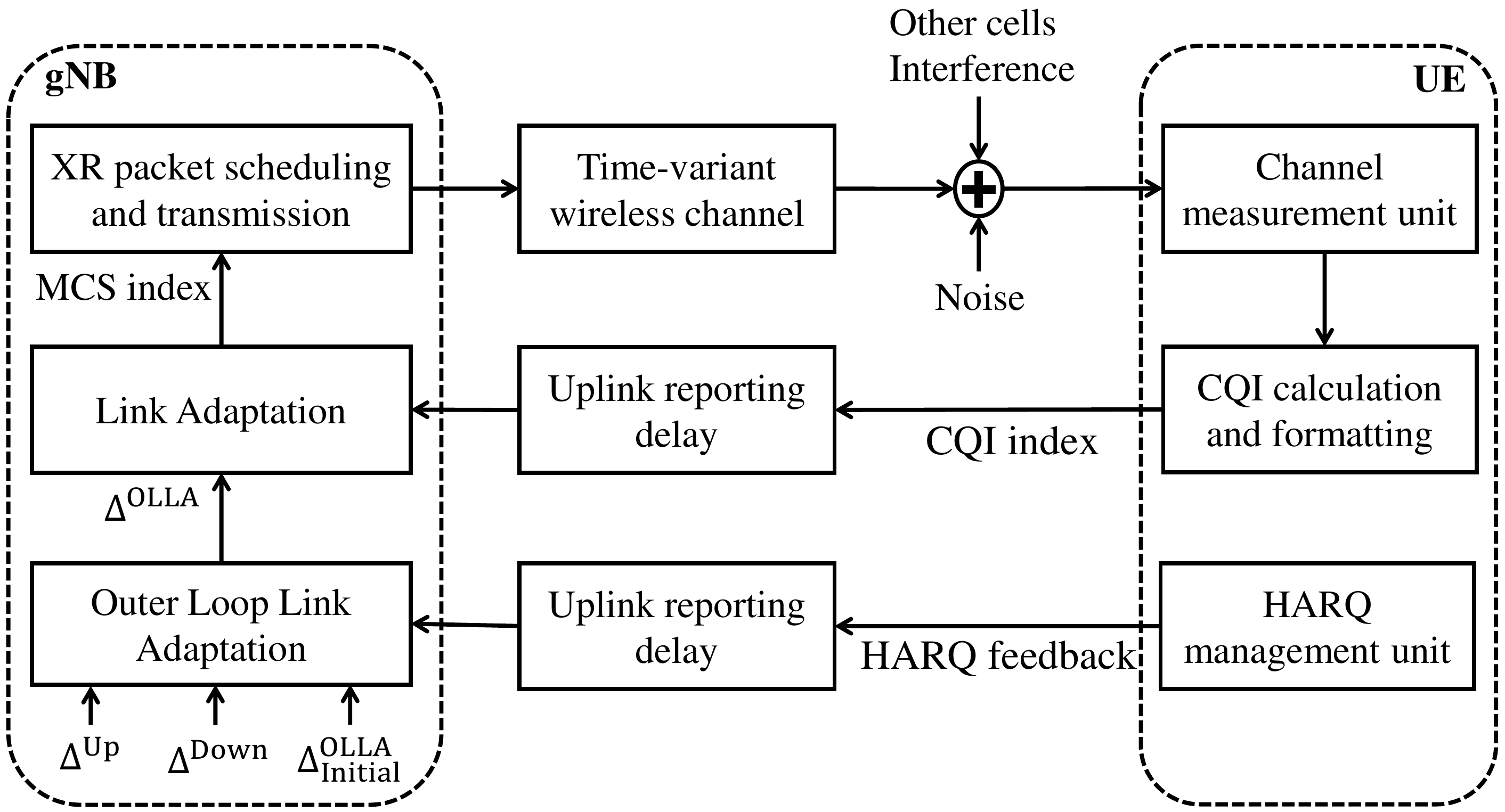}}
    		\caption{Simplified block diagram showing the link adaptation mechanism and the partial feedback information from the UEs.}
    		\label{P2:Fig:OLLA_CQI_Tx_Rx}
\end{figure}
 The overall framework for dynamic link adaptation (LA) is illustrated in Fig. \ref{P2:Fig:OLLA_CQI_Tx_Rx}. The gNB relies on the CQI feedback from the UEs to select the MCS for its downlink transmissions. The currently standardized 5G NR CQI scheme corresponds to the highest supported MCS index that the UE can decode with a transport block error probability (TBEP) of less than a certain predefined value (default is 10\%). We refer to this as the baseline CQI. Note that the baseline CQI is not having any explicit relation with CBG-based transmissions as it was originally designed for the TB-based transmission with single-bit HARQ feedback. The UE can be configured to report the CQI with a certain periodicity over the number of frequency sub-bands $ \mathcal{S} = \{1, \cdots, S\}$. Reporting formats with \textbf{wideband} ($|\mathcal{S}| = 1$) and \textbf{sub-band}($|\mathcal{S}| > 1$)are supported \cite[Table 5.2.2.1-1]{3gpp.38.214}. The UE obtains the CQI by performing measurements of the received post-detection SINR. It uses the measured SINR to look up the highest supported MCS with a maximum TBEP of less than 10\%, assuming the UE has tables with its TBEP performance versus SINR for different MCSs. See more details for the 5G NR CQI reporting format in \cite[Sec. 5.2]{3gpp.38.214}.

 As illustrated in Fig. \ref{P2:Fig:OLLA_CQI_Tx_Rx}, the control loop for link adaptation includes time-variant radio channels, interference, and also delays in CQI reporting and potential UE measurement imperfections for obtaining the CQI. Therefore, it is common practice to also apply an OLLA scheme. The OLLA scheme is at the gNB side, where it monitors the success of past downlink transmissions through the HARQ feedback and imposes some additional offset on the received CQI before applying it for MCS selection. Traditional OLLA for TB-based transmission with Boolean (ACK or NACK) HARQ feedback is studied in  \cite{blanquez2016eolla}, and many other references. An enhanced OLLA (eOLLA) was recently explored in \cite{GlobeCom_eOLLA} that is tailored to CBG-based transmissions where multi-bit HARQ feedback is coming from the UE, i.e. ACK/NACK per CBG. As is described in these OLLA/eOLLA references, the gNB is able to more accurately control the experienced TBEP/CBGEP through proper parameterization of step-up and step-down sizes of the so-called OLLA offset that is applied to the CQI before using it for MCS selection.

\subsection{XR KPIs}	
We adopt the XR capacity definitions as earlier agreed by 3GPP in  \cite{3gpp.38.838}. Here an XR UE is marked as satisfied if the UE successfully receives more than X\% of the packets within a given packet delay budget (PDB). The XR Capacity is the maximum number of XR users that can be supported per cell with at least Y\% of them being satisfied. It is assumed that X=99\% and Y=90\%. Our overall goal is essentially to propose eCQI enhancements that lead to higher XR capacity as compared to baseline cases with current CQI methods. In addition to the XR user satisfaction and capacity, we also consider more generic radio KPIs in our study. Those include the radio resource utilization that is measured by monitoring the PRB utilization of the gNBs. The latter is a useful measure to assess if our proposals result in higher radio resource efficiency for certain offered traffic conditions. In addition, latency performance per XR packet and error probabilities for HARQ transmissions are considered.

\section{Analytical Assessment of CBG-based Transmissions}\label{Analytical_Assessment}
In the following, the behavior of CBG-base transmissions is further analyzed as the first step toward setting the scene for deriving eCQI schemes in such cases. 
We examine the TB and CBG error probability behaviors for different cases, including correlated and non-correlated error events.
\subsection{Identical and Independent CBGs Case}

    We start by defining the error probability of a TB, CB, and CBG as $p^r_{\text{TB}}$, $p^r_{\text{CB}}$, and $p^r_{\text{CBG}}$, respectively, for the special case of i.i.d. errors of CBs. Here $r$ is the MCS index as defined in \cite[Table 5.2.2.1-2-5]{3gpp.38.214}. A corresponding analysis under such i.i.d. conditions is available in  \cite{khosravirad2017flexible, Reduced_CBG_HARQ_for_coexistence_with_urllc_traffic, yeo2017partial}. Therefore, $p^r_{\text{CB},i}=p^r_{\text{CB},j}=p^r_{\text{CB}}$, $\forall i,j \in \{1,\cdots,C\}$.    In order to find the CBGEP, for each CBG $m$, we use the error probability of the CBs within that CBG with the set of indices $\mathcal{A}_m$, which leads to
    \begin{align}\label{eq: CBG error probability - iid case}
       p_m^{\text{CBG},r} = 1-(1-p_i^{\text{CB},r})^{|\mathcal{A}_m|}.
    \end{align}
If all CBGs contain the same number of CBs, CBGEP will be the same for all CBGs \cite{Preemptive_CBlayouts} and \eqref{eq: CBG error probability - iid case} can be rewritten as
     \begin{align}\label{eq: CBG error probability - iid case2}
       p^r_{\text{CBG}} = 1-(1-p^r_{\text{CB}})^{C/M},
    \end{align}
    where subscripts $m$ and $i$ are omitted due to identical properties of the CBs and the CBGs.
    Finally, the TBEP can be expressed as 
    \begin{equation}\label{P2:Eq:accurate_and_approx_TBLER}
    		p^r_{\text{TB}}=  1-\left(1-p^r_{\text{CBG}}\right)^M = 1-\left(1-p^r_{\text{CB}}\right)^C.
    \end{equation}
    Using simple substitutions, we rewrite the CBGEP as:
    	\begin{equation}\label{P1:Eq:CBGER}
    		p^r_{\text{CBG}}=1-\left(1-p^r_{\text{TB}}\right)^{\frac{1}{M}}.
    	\end{equation}
    Assuming that $n_e$ denotes the number of failed CBGs in a TB, the probability of having at most $N$ failed CBGs out of $M$ CBGs can be obtained as \cite{GlobeCom_eOLLA}:
    \begin{align}\label{P1:Eq:EP_num_failed_CB}
    		P_e(r,n_e \leq N) = \!\!
    		\sum_{n_e=0}^{N}\!\!
    		\binom{M}{n_e} \left(p^r_{\text{CBG}}\right)^{n_e} \left(1-p^r_{\text{CBG}}\right)^{M-n_e}.        
    \end{align}
\subsection{Non-identical and Non-correlated Case}

We next investigate the cases where the i.i.d. assumptions are no longer enforced. That is, considering cases with correlated errors among CBs, and also having different error probabilities per CB (and CBG). For such cases, the CBGEP can be expressed as
    \begin{align}\label{eq: CBG error probability}
       p_m^{\text{CBG},r} = 1-\bigcap_{i\in \mathcal{A}_m} (1-p_i)^{\text{CB},r},
    \end{align}
    that generalizes \eqref{eq: CBG error probability - iid case} for uneven CB error probabilities which can be a result of variable effective SINR among the CBs of one CBG. 
    For the sake of simplicity, we now drop the superscript CBG and $p^r_m$ will refer to CBGEP of CBG $m$ at MCS index $r$.
    The probability of only one CBG failing i.e, $N=1$, is expressed:
    \begin{align}\label{eq: N eq 1 case}
        P_e (r,1)  = \sum_{m \in \mathcal{M}} p^r_m \bigcap_{j \in \mathcal{M}\setminus m}(1-p^r_j),
    \end{align}
     where $\mathcal{M}=\{1,\cdots,M\}$ and $\mathcal{M}\setminus m$ refers to set exclusion where element $m$ is discarded from $\mathcal{M}$. Extending this expression for a general value for $N$ results in:
 
    \begin{equation}\label{eq: N general case}
        \begin{aligned}
        P_e (r,N) &= \frac{1}{N!}\underbrace{\sum_{m_1 \in \mathcal{M}^{(0)}} p^r_{m_1} \cdots \sum_{m_N \in \mathcal{M}^{(N-1)}} p^r_{m_N}}_{N \text{times}}
        \quad \\&\times\bigcap_{m_{N+1} \in \mathcal{M}^{(N)}}(1-p^r_{m_{N+1}}),
        \end{aligned}
    \end{equation}
    which is the direct extension of Binomial distribution in \eqref{P1:Eq:EP_num_failed_CB}. Here, $\mathcal{M}^{(t)}$ is set $\mathcal{M}$ with the exclusion of $t$ elements where this exclusion is done after each of the inner summations and product operators (for \eqref{eq: N eq 1 case} $\mathcal{M}\!\setminus\! m$ is equivalent to the case  $t=1$). As can be seen from \eqref{eq: N general case}, as $N$ grows, these calculations become tedious to evaluate. We return to this issue in \ref{subsection: closed-form eq} and propose some methods for complexity reduction.

\subsection{Correlated Case}
    So far, we assumed independence between $p_m$s which may not always be the case in reality. So here we study the correlated scenario for identical $p_m$s. Assume a probability distribution of $Y= W_1 +\cdots + W_M$, where $W_m$, $m=1,\cdots, M$ are binary variables (i.e. CBG CRC checks) with mean value and variance of $E(W_m)=p_m$ and $\text{Var}(W_m)=p_m(1-p_m)$, respectively. Additionally, there is a correlation coefficient of $\text{Corr}(W_m,W_n)=\rho$, $m\neq n$. In our model, $0\leq\rho\leq 1$ is a function of similarities between the assigned PRBs to each of the CBGs. The probability distribution of $Y$ is found by the mixture of the distributions of two variables \cite{diniz2010bayesian}: a binomial distribution, $B(N, p)$, where $p=p_1=\cdots=p_m$ is CBGEP, with a mixing coefficient of $(1-\rho)$ and a modified Bernoulli distribution, $MBern(p)$\footnote{This distribution is taking values $0$ or $n$ \cite{fu1995generalization}, rather than the
    conventional values 0 or 1.} with a mixing probability of $\rho$. Thus, the probability distribution
    of $Y$ is given by
    \begin{equation}
        \begin{aligned}
            \hspace{-0.12 cm}\Tilde{P}_e(r,N,\rho,M) \!&=\!\! {\binom{M}{N}} \!(p^{r})^N (1\!-\!p^r)^{M-N}(1\!-\!\rho)\delta_{A_1}\!(\!N)
            \\& + p^{r,\frac{N}{M}} (1-p^r)^\frac{M-N}{M}\rho \delta_{A_2}(N)
        \end{aligned}
    \end{equation}
    where $p^r=p_1^r=\cdots=p^r_m$ is CBGEP for all CBGs for MCS $r$, $A_1=\{0,1,\cdots,M\}$, $A_2=\{0,M\}$ and $\delta_\mathcal{X}(x)$ is the Delta function that is equal to $1$ if $x\in\mathcal{X}$ and is $0$ otherwise. It is worth mentioning that we do not study the case of correlated and non-identical probability function, i.e. generalized form of \eqref{eq: N general case}, as it is very complex to derive a closed-form function.
    
	\begin{figure}[t]
		\centerline{\includegraphics[width=0.65\textwidth,trim={0cm 0.0cm 0cm 1.15cm},clip]{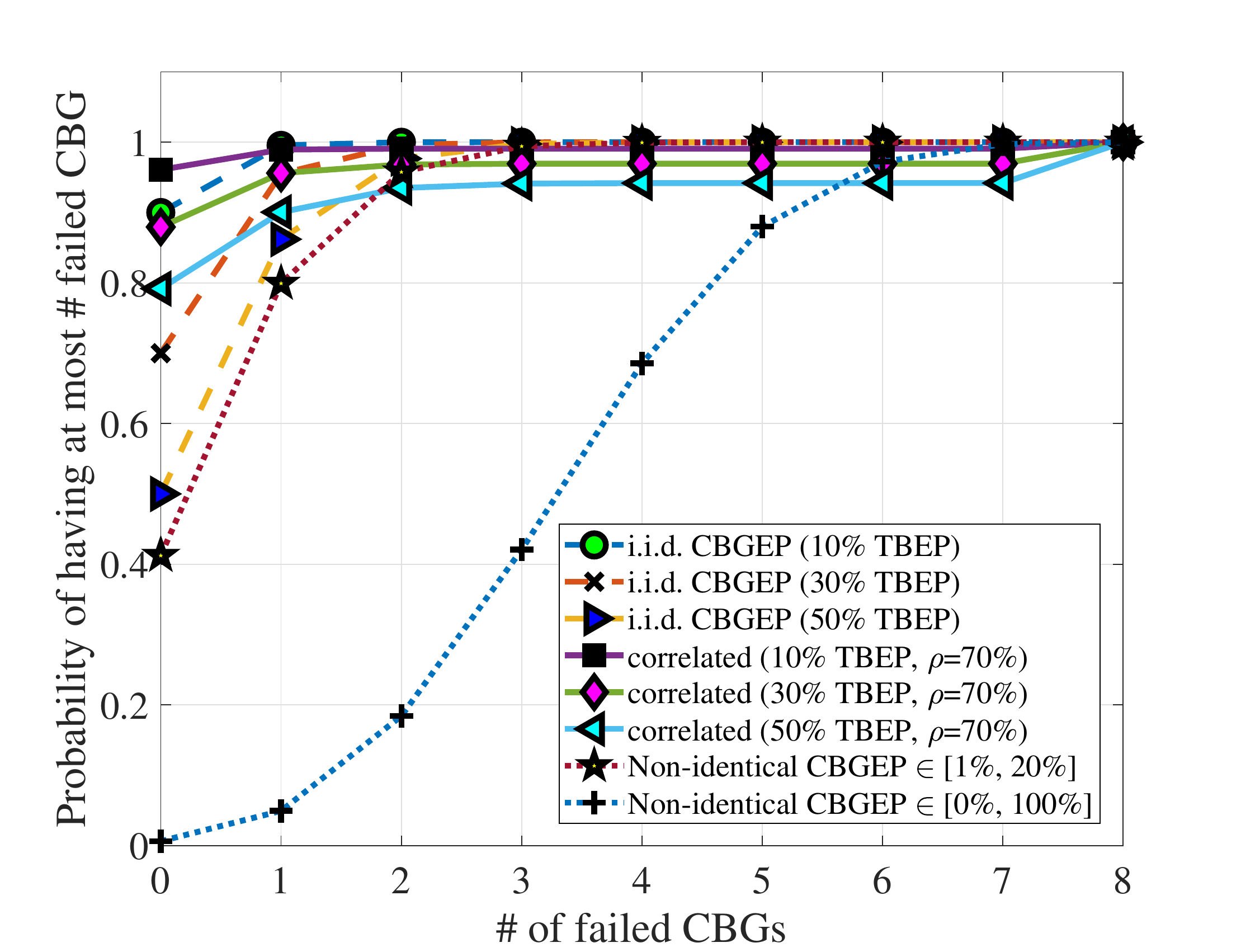}}
		\caption{The probability of having at most $N$ failed CBGs for identical and non-identical cases when there are $8$ CBGs in a TB.}
		\label{P1:Fig:failedCBGs_BLER}
	\end{figure}
	In Fig. \ref{P1:Fig:failedCBGs_BLER}, we show the probability of having at most $k$ failed CBGs in a TB for different identical, correlated, and non-identical cases.
	For correlated cases, we assumed a correlation coefficient of 70\% which is considered to be a strong correlation between CBGEPs. As can be seen, the probability of having two or less failed CBGs is roughly close to 1 for i.i.d cases (as an example, the ratio is \SI{99.8}{\percent} if $P_{e_\text{TB}}=\SI{10}{\percent}$.).  It is observed that even for the correlated or non-identical CBGEP cases there is a high probability of 90\% having two or fewer failed CBGs in a TB. The worst-case scenario in the figure happens when CBGs experience non-identical random CBGEP distributions between 0 to 1. As an example of the  worst-case scenario, the probability of having at most five failed CBGs is about 85\%. The results also clearly show that if using the current baseline CQI scheme with a fixed TB error rate target of 10\%, the CBG errors will vary significantly depending on the assumptions. This is undesirable as we prefer a link adaption scheme where we more accurately control the number of failed CBGs per TB transmission, and the corresponding error probabilities. This is desirable, so we are able to control the number of resources used on HARQ retransmissions as this is important for maximizing the XR capacity.

\section{eCQI for CBG-based transmissions}\label{CQI enhancements for CBG}
Our hypothesis is that we will be able to further boost the XR capacity by introducing an eCQI scheme where we are able to more explicitly control the CBG errors, as compared to using the current baseline CQI that aims for 10\% TBEP. What we propose is an eCQI scheme where at most $N$ out of $M$ CBGs in a TB transmission are in error with probability $P$. This allows controlling the radio resources used for HARQ retransmissions while using the highest possible MCS under such constraints. Given this starting, we present an eCQI solution for this, assuming that parameters $N$, $M$, and $P$ are configured by the network for the UEs; may e.g. be realized by RRC signaling as part of the channel state information (CSI)-ReportConfig information element as defined in \cite{3gpp.38.331}.

The proposed eCQI method is summarized in  Alg.~\ref{P2:Alg:CBG-based_eCQI}.
    \begin{algorithm}[t]
	\SetAlgoLined
	\KwResult{Highest MCS index to satisfy a given condition}
	\emph{Initialize:} 
	 $M$,  $N$, $P$, tables to convert SINR (or MMIB) to error probability per each MCS index. 
	 
    1. Calculate the effective SINR of each CB $\gamma_i^{\text{CB}}, \forall i \in \{1,...,C\}$. 
    
    \For{$r \in \mathcal{R} $ }{

    2. Calculate error probability of each CB $p_i^{\text{CB}}, \forall i \in \{1,...,C\}$ using curves for MCS index of $r$.
    
    3. Find CBG error probability $p_m^r$ and $m \in \{1,...,M\}$ using \eqref{eq: CBG error probability}.
    
    4. Calculate $P_e(r,N)$ using \eqref{eq: N general case}.
    
    \uIf{{Stopping criterion is met (e.g. for eCQI $\hat{P_e}(r,N)>P$)}}
    {
    5. Break.
    }
    \uElse{}{
    6. Choose the next element in $\mathcal{R}$ for $r$.
    }
    }
    
    7. report $r\!-\!1$ as the best MCS index.
    
    	\caption{Generic calculations for eCQI.}
    	\label{P2:Alg:CBG-based_eCQI}
    \end{algorithm}    
    After the UE has performed measurements of its experienced SINR, the effective SINR of each CB can be calculated. The effective SINR values are mapped to error probability values for each MCS index $r$ employing tables in the UE that are pre-configured \cite{Reduced_CBG_HARQ_for_coexistence_with_urllc_traffic}. This may also be conducted using the mean mutual information per bit (MMIB) metric and then using corresponding tables to determine error probabilities \cite{tang2010mean}. Thereafter, $p_i^{\text{CB},r}$s are used to calculate CBGEP from \eqref{eq: CBG error probability} and $P_e(r,N)$ from \eqref{eq: N general case}. These steps are repeated for the MCS indices $r \in \mathcal{R}$  until the stopping condition of $\hat{P}_e(r,N)=\sum_{i=0}^N P_e(r^\ast,i)>P$ is met. Finally, $r^\ast-1$ is reported as the highest supported eCQI index. Similar to the baseline CQI scheme, eCQI reporting may be periodical or event-based. 

\section{Complexity reduction for eCQI calculation} \label{Computational_Complexity_Reduction}
The eCQI should be realized in the UE with as low complexity as possible. As can be seen from \eqref{eq: N general case}, as $N$ grows, these calculations become tedious to evaluate and the problem becomes more severe due to limited processing power. In order to address this challenge, we derive closed-form expressions for $P_e(r,N)$ to reduce the eCQI complexity. These include several attractive searching techniques for the main loop in Alg. \ref{P2:Alg:CBG-based_eCQI}.

\subsection{Closed-form Solutions for Special Cases} \label{subsection: closed-form eq}
Considering an arbitrary number of CBGs in one TB $M\leq 8$, we derive closed-form expressions for the probability calculations in \eqref{eq: N general case} for values of $N \in \{0,1,2,3\}$. Expressions for $N>3$ are more complex and omitted for this paper. However, the extension of the methods used for $N<4$ cases is straightforward for any arbitrary $N$ and can be achieved by following the same steps as the cases stated below.
Before starting to derive the closed-form solutions, we define the \textit{Odds ratio (OR)} of a random variable which is defined as
    \begin{align}\label{eq: OR transform}
        O_m=\frac{p_m}{1-p_m}
    \end{align}
which shows the ratio between the success and failure probability of an event for our Bernoulli random variables. Now, by rewriting \eqref{eq: N eq 1 case}, we have
    \begin{equation}\label{eq: N eq 1 case - proposed}
        \begin{aligned}
            P_e (r,1)  &= \sum_{m \in \mathcal{M}} p^r_m \bigcap_{j \in \mathcal{M}\setminus m}(1-p^r_j) \\&= \Pi\sum_{m \in \mathcal{M}} \frac{p^r_m}{1-p^r_m} = \Pi \sum_{m \in \mathcal{M}} O^r_m = \Pi M \Bar{O^r}
        \end{aligned}
    \end{equation}
    where, $\Pi=P_e (r,0)=\bigcap_{m \in \mathcal{M}}(1-p^r_m)=\bigcap_{m \in \mathcal{M}}\frac{1}{1+O^r_m}$ and the bar operator in $\Bar{X}$ stands for the average value of variable $X$.
     
    Extending this result for the case of $N=2$ is straightforward. Using \eqref{eq: N general case} for $N=2$ leads to
    \begin{equation}
        \begin{aligned}\label{eq: N eq 2 case}
            P_e (r,2)  \!&=\!\frac{1}{2!}\!\sum_{m_1 \in \mathcal{M}^{(0)}} \!\!p^r_{m_1}\!\!\sum_{m_2 \in \mathcal{M}^{(1)}}\!\!\!p^r_{m_2} \quad \!\!\!\!\!\!\!\!\!\!\bigcap_{m_{3} \in \mathcal{M}^{(2)}}\!\!\!\Bigl(1-p^r_{m_{3}}\Bigr)
            \\&= \frac{\Pi M^2}{2}\Bigl(2 \Bar{O^r}^2 - \text{Var}\bigl(O^r\bigr)\Bigr)
        \end{aligned}
    \end{equation}
    where $\text{Var}(\cdot)$ is the variance operator. 
    \begin{proof}
	Please see Appendix \ref{Appendix A}.
	\end{proof}
      
    Finally, extending the solutions for the case with $N=3$ with a similar approach is given by.
        \begin{equation}\label{eq: N eq 3 case}
            \begin{aligned}
             &P_e (r,3) = \frac{\Pi}{3!} \!\!\!\sum_{m_1 \in \mathcal{M}} \!\!\!O^r_{m_1} \!\!\!\sum_{m_2 \in \mathcal{M}\setminus m_1} \!\!\!O^r_{m_2} \!\!\!\sum_{m_3 \in \mathcal{M}\setminus m_1,m_2} \!\!\!\!\!\!O^r_{m_3}
            \\&= \frac{M^3\Pi}{6}\Biggl[-2\Bigl(\Bar{O^r}\Bigr)^3 \!-\! 3 \Bar{O^r}\text{Var}\Bigl(O^r\Bigr)  \!+\! 2 \!\!\sum_{m \in \mathcal{M}} \!\!\Bigl(O^r_{m}\Bigr)^3\Biggr]
            \end{aligned}
        \end{equation}
    \begin{proof}
	Please see Appendix \ref{Appendix B}.
	\end{proof}
    Note that the last term in \eqref{eq: N eq 3 case} can be calculated using the 3rd derivation of the moment generating function of the random variable $O^r$. Further cases for $N>3$ closed-form solutions can be obtained by similar approaches used so far. The key point is to start from \eqref{eq: N general case} and convert the expressions to be a function of the OR. Then, by factorizing similar summations, low complexity expressions can be found directly.
     
     If the eCQI is configured in a way that it ensures having \textbf{exactly} $N$ failed CBGs out of $M$ CBGs with probability $P$, then the expressions derived in \eqref{eq: N eq 1 case - proposed}, \eqref{eq: N eq 2 case}, and \eqref{eq: N eq 3 case} can be used directly. For instance, if the configuration notifies at exactly $2$ failed CBGs, then the threshold comparison should be done with $P_e(r,2)$.

\subsection{CQI Index Searching Schemes}
    Another main component in controlling the complexity of the eCQI is the searching method of different MCS indices. In other words, if the MCS index list $\mathcal{R}$ is ordered in a smart way, a proper eCQI (or CQI) index can be obtained faster with fewer computations. Therefore, we analyze and propose different methods to optimize the searching space size to adjust the accuracy/complexity trade-off to find the best MCS index.
 
\subsubsection{Descending / Ascending Order} \label{subsection: Descending/ascending order}
    This linear search method (LSM) is the default way of iterating Alg. \ref{P2:Alg:CBG-based_eCQI} over different MCSs.
    For the case of descending order, the ordered list is made as the following $\mathcal{R}=\{r_1,\cdots,r_I|r_i>r_{i+1}\}$ with $I$ as the total number of MCS indices. Similarly, for the case of ascending order, we have $\mathcal{R}=\{r_1,\cdots,r_I|r_i<r_{i+1}\}$. 
    The benefit of this type of listing is that they have almost no overhead complexity for generation at the receiver. On the other hand, they are not optimal and may result in added processing time since Alg. \ref{P2:Alg:CBG-based_eCQI} may require iterating over many elements before reaching the right index. In the following, we present several alternative searching methods to speed up the eCQI calculations.

\subsubsection{Binary Search Method (BSM)}
\begin{figure}[t]
    		\centerline{\includegraphics[width=0.65\textwidth]{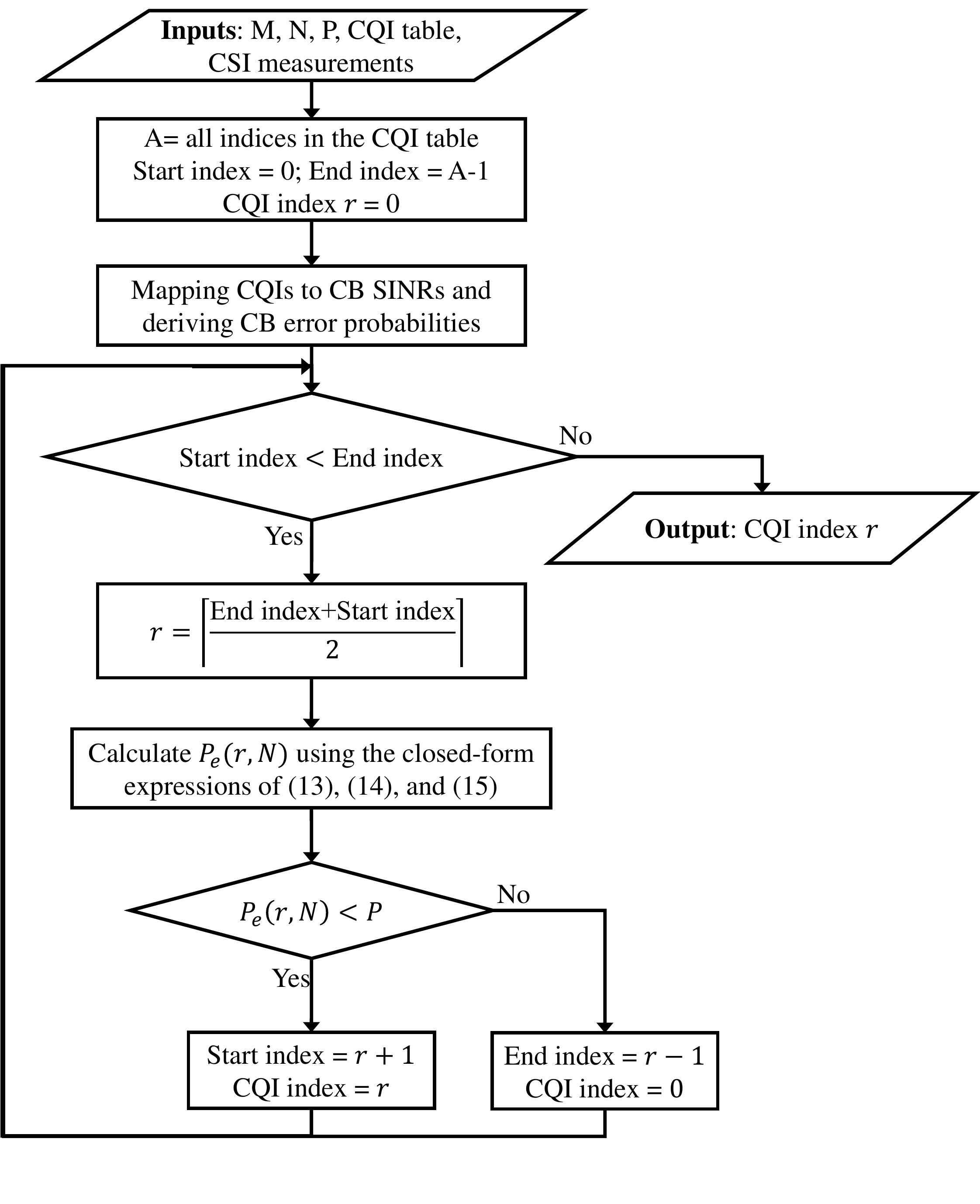}}
    		\caption{The flowchart of embedded BSM in the eCQI index determination.}
    		\label{P2:Fig:Flowchart_BSM_eCQI}
    \end{figure}

    This method focuses on the order of steps in which the MCS index is checked in Alg.\ref{P2:Alg:CBG-based_eCQI}. Therefore, a simple index list, such as the one described in subsection \ref{subsection: Descending/ascending order}, can be used to begin with.  Then, $r=I/2$ can be used for the initial value. After calculating $P_e(r,N)$, depending on its comparative logic to $P$, the next $r$ will be decided using a simple BSM. In other words, starting with $r_{\frac{I}{2}}$ and for each $r$, if $P_e(r,N)>P$, then the next index should be in the lower half of the MCS list. Thus, the next $r$ is $r_{\frac{I}{2}-\frac{I}{4}}=r_{\frac{I}{4}}$. Likewise if $P_e(r,N)<P$, then the next index should be in the upper half of the MCS list and $r_{\frac{3I}{4}}$. This mechanism continues until for some $r_i$, we have  $P_e(r_i,N)<P$ and $P_e(r_{i+1},N)>P$. A flowchart of embedded BSM in the eCQI index determination is shown in Fig. \ref{P2:Fig:Flowchart_BSM_eCQI}. As illustrated there, for each entry in the CQI table, the UE calculates \eqref{eq: N general case} using one of the closed-form solutions provided in the previous section. Depending on the outcome of $P_e(r,N)\overset{?}<P$, the next index is determined via the BSM. The process continues until the highest index that satisfies the target probability condition is found.

\subsubsection{Relaxed Decision Criterion } 

    This method is an alternative to the argument in the `if' condition in Alg.\ref{P2:Alg:CBG-based_eCQI}, where instead of having a firm decision boundary at $P$,  a small tolerance factor is included in the condition. This can be done with $|P_e(r,N)-\delta_p|\overset{?}{<}P$, where $\delta_p$ allows a small room for the error probability variation. This  condition can be used in two ways: 1) any $r$ satisfying it is reported or 2) the next $r$ in $\mathcal{R}$ is reported without going into the steps of Alg.\ref{P2:Alg:CBG-based_eCQI} to avoid additional calculations. 
 
    In order to choose a proper searching method, we evaluate the complexity of each method in the next subsection.  
 
\subsection{Complexity Analysis}
    We compare the computational complexity of the direct calculations of \eqref{eq: N general case} and the closed-form solutions for the eCQI technique in terms of the number of multiplications. 
    We focus on two main parts which are the execution of \eqref{eq: N general case} and the size of the CQI index search space $|\mathcal{R}|$. The reason for this choice is that these two are the main components of controlling the complexity of Alg.\ref{P2:Alg:CBG-based_eCQI}, as the former one is repeated $1\leq t \leq |\mathcal{R}|$ times until the algorithm finds the proper CQI index. It is true that there are several other steps in Alg.\ref{P2:Alg:CBG-based_eCQI} that require processing resources, yet for the sake of comparison, we ignore those since they have the same complexity no matter which method is used for probability derivations.
 
\subsubsection{Complexity for the Execution of \eqref{eq: N general case}}
    As a baseline method, we assume direct calculations of \eqref{eq: N general case}, where a series of multiplications are done sequentially until all cases of failed CBGs are covered based on a given $N$. Starting from the simplest case of $N=1$ in \eqref{eq: N eq 1 case}, the complexity is equal to $M (M-1) =M^2-M=\mathcal{O}(M^2)$ multiplications. For $P_e(r,2)$ the result is $M (M-1)M/2 =\mathcal{O}(\frac{1}{2}M^3)$. With a straightforward extension for any arbitrary $N$, the complexity for the execution of \eqref{eq: N general case} is given by
    \begin{align}
         C_{\text{Direct}} = M\binom{M}{N}=\frac{MM!}{(M-N)!N!}=\mathcal{O}(\frac{1}{N!}M^{N+1}).  
    \end{align}

    With a similar approach, we try to derive the complexity of our proposed method to calculate \eqref{eq: N general case}. It is worth mentioning that the proposed expressions are in form of functions of the OR $O^r$ and thus, it is expected that calculating $O^r$ is done directly from SINR values. This can be done utilizing a set of new curves that map SINR to $O^r$ and can be generated using the SINR to TBEP curves and  transformation in \eqref{eq: OR transform}.
    For the case of $N=1$ in \eqref{eq: N eq 1 case - proposed}, the only term containing multiplications is $\Pi$ that requires $M=\mathcal{O}(M)$ multiplications. Similarly, for $N=2$ in \eqref{eq: N eq 2 case} we have $M+M+1=2M+1=\mathcal{O}(2M)$ multiplications which is from calculating $\Pi$, $\sum_{m_1 \in \mathcal{M}} (O^r_{m_1})^2 $ and $(\sum_{m_1 \in \mathcal{M}} O^r_{m_1})^2$, respectively.  Likewise, for $N=3$ in \eqref{eq: N eq 3 case} there are $3M+3=\mathcal{O}(3M)$ multiplications. Thus, by extending these results, for a general $N$ value we have 
    \begin{align}
    C_{\text{Closed}} = \mathcal{O}(NM).
    \end{align}
    As can be seen from the complexity expressions, our proposed method has a better behavior concerning $M$ and $N$, i.e. $C_{\text{Direct}}>C_{\text{Closed}}$. The lower complexity is a result of factorizing the repeated operations and doing them only once instead of calculating long summation loops. 
 
\subsubsection{CQI Index Searching Space Size}
    Another main component in determining the complexity of Alg.\ref{P2:Alg:CBG-based_eCQI} is the number of iterations after step 1 that scales the number of total multiplications linearly. 
    For the baseline case of descending/ascending order, the size of the search space is $|\mathcal{R}|$ in the worst-case scenario and is equal to $\frac{|\mathcal{R}|}{2}$ on average. This is due to the fact that the UE will start checking all indices from the beginning to the end of the CQI tables of \cite[Table 5.2.2.1-2-5]{3gpp.38.214} and on average it may end up searching half of the entries. On the other hand, the complexity of the BSM is much lower and is $\log_2(|\mathcal{R}|)$ due to the step-wise halving of the search space. Finally, for the relaxed decision scheme, the search space size could be realized as $\alpha \log_2(|\mathcal{R}|) $ with $0<\alpha=f(\delta_p)\leq 1$, where the complexity can be scaled as a function of $\delta_p$. However, there is an inverse relation between $\alpha$ and TBEP, as for a more relaxed decision criterion, the resulting error rate will increase.
    \begin{figure}[t]
    		\centerline{\includegraphics[width=0.55\textwidth]{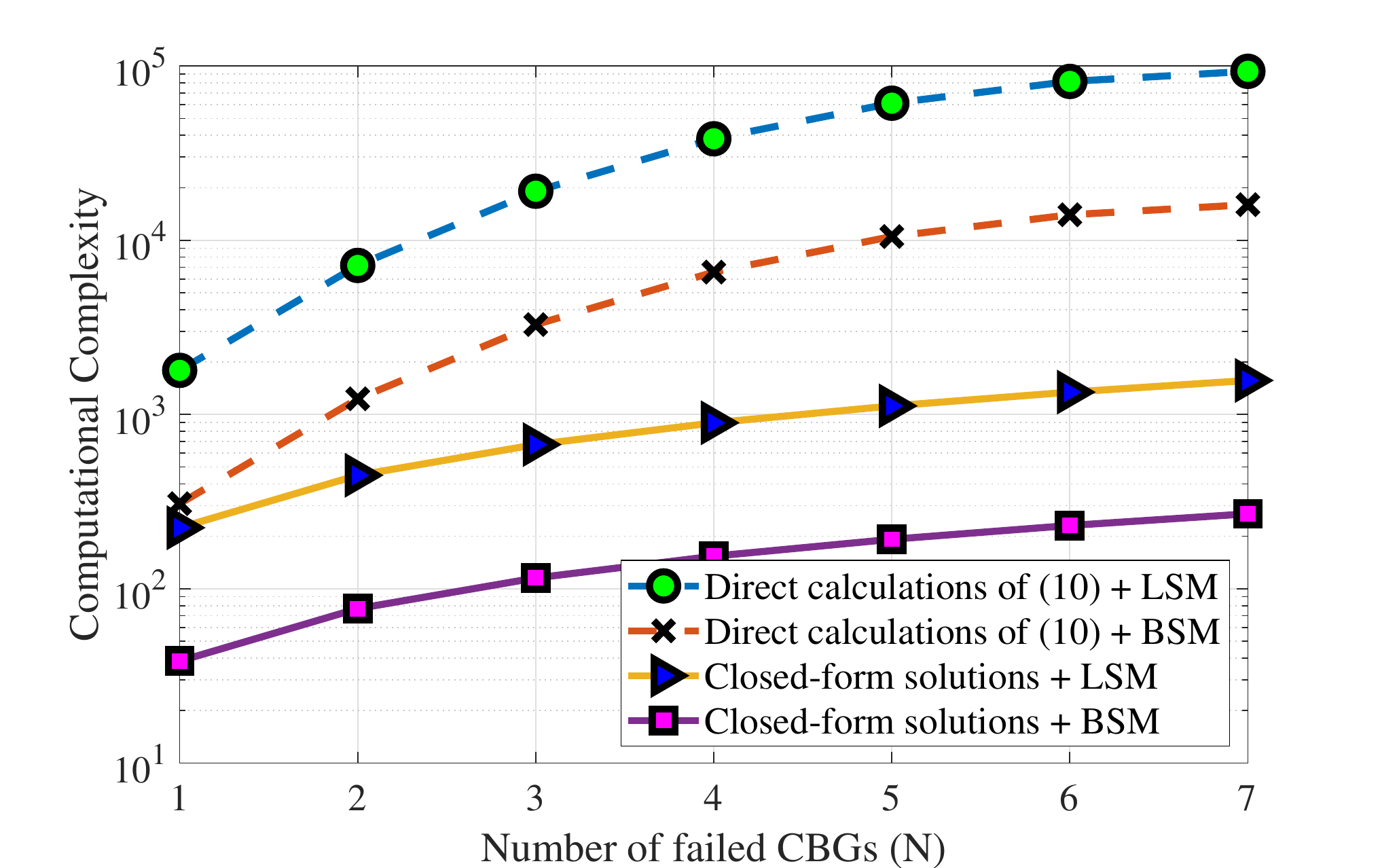}}
    		\caption{Complexity comparison between the proposed and baseline methods in terms of number of multiplications in logarithmic scale.}
    		\label{P2:Fig:Complexity}
    \end{figure}
    Figure \ref{P2:Fig:Complexity} shows a comparison of the complexity of different combinations of eCQI determination and searching methods. As can be seen, the lowest computational load and our preferred solution can be achieved by choice of the closed form type solution and a BSM for the CQI index selection part.

\section{Numerical Evaluations}\label{Simulation_Results}
\subsection{3GPP-Specific Simulation Methodology}

    To evaluate the performance of the proposed methods, we use highly realistic dynamic system-level simulations, where the performance-determining parts of  radio access network mechanisms such as dynamic packet scheduling, link adaptation, HARQ,  as well as time- and frequency-varying inter-cell interference are accurately accounted for. We adopt the simulation methodology and modeling for 5G NR, as agreed in the 3GPP technical report \cite{3gpp.38.838}, and also used in the following studies \cite{petrov2022extended, GlobeCom_eOLLA}. We chose system-level simulations as our methodology here to obtain results with a high degree of realism in highly dynamic multi-cell and multi-user settings as deriving the performance in terms of achievable XR capacity is not analytically traceable.  All simulations are carried out using the Monte Carlo approach with long simulation runs to obtain statistically reliable results. 

    \begin{table}[t]
		\centering
		\caption{Summary of System-level Evaluation Parameters}
		\begin{tabular}{cc|cc}
			\textbf{Parameter} & \textbf{Setting} &\textbf{Parameter} & \textbf{Setting} \\  \toprule 
			Simulation time & 10 seconds (20,000 slots) & Simulation runs  & 10 runs per result \\ 
			Deployment & InH (120m$\times$50m) & Layout  & 12 cells  \\ 
			Inter-site Distance & 20 m	& TDD Frame structure & DDDSU \\ 
			TTI length & 14 OFDM symbols & Carrier frequency & 4 GHz \\ 
			System Bandwidth & 100 MHz & SCS & 30 kHz \\ 
			MIMO scheme & SU-MIMO with rank 1  & Modulation & QPSK up to 256QAM \\
			gNB height & 3 m  & gNB power & 31 dBm \\ 
			gNB Tx processing delay & 2.75 OFDM symbols & \multirow{2}*{gNB antenna} & 1 panel with 32 elements \\
			& & & (4 × 4 and 2 polarization) \\ 
			UE Rx processing delay & 6 OFDM symbols  & UE height & 1.5 m  \\ 
			UE speed & 3 km/h & \multirow{2}*{UE antenna} & 4 omni-directional antennas \\
			& & & (2 × 1 and 2 polarization) \\
			Jitter distribution & $\mathcal{TN}(0, 2, -4, 4)$ ms & Frame size (30Mbps) & $\mathcal{TN}(62, 6, 31, 93)$ kB \\
			Frame size (45Mbps) & $\mathcal{TN}(93, 10, 46, 140)$ kB & Frame rate & 60 fps \\
			Source data rate & 30 Mbps, 45 Mbps & Packet delay budget & 10 ms, 15 ms \\ 
			HARQ scheme & Chase combining & HARQ retransmissions &  Up to 3 retransmissions\\ 
			CSI & Periodic CQI every 2 ms  & Initial OLLA Offset & 0 dB \\ 
			Upper limit of OLLA Offset & 15 dB & Lower limit of OLLA Offset & -25 dB \\ \toprule
		\end{tabular}
		\label{P2:Table:System_Parameters}
	\end{table}
    Simulation assumptions/parameters are summarized in Table \ref{P2:Table:System_Parameters}. 
    We simulate an InH scenario as described in \cite{3gpp.38.838}, where the 3D radio channel propagation model is calibrated against alike results published in 3GPP.  The scenario is composed of 12 gNBs. XR UEs are randomly distributed in the network according to a spatial uniform distribution. Serving cell selection is based on the strongest reference signal received power (RSRP) criteria. TDD mode is adopted, where we assume a DDDSU slot pattern. Each slot is having a duration of 0.5 ms and is composed of 14 OFDM symbols where the first symbol of each slot is always carrying control overhead such as scheduling grants on the physical downlink control channel (PDCCH).  In the special slot, the first ten symbols are allocated for downlink, two symbols are for a gap, and the last two symbols are for uplink. Simulations are conducted at the symbol level (time-domain) and sub-carrier resolution (frequency-domain).
    
    For LA purposes, UEs report the CQI index periodically in the uplink slot. We assume CQI reporting every 2 ms with a CQI reporting delay of 2 ms.  The scheduler is PF. Asynchronous adaptive HARQ with Chase Combining for failed transmissions is assumed, where the post-processing SINR equals the linear sum of SINRs of the different HARQ transmissions (first transmission and subsequent retransmissions) \cite{chase1985}. Error-free ACK/NACK feedback is assumed from UEs. The maximum number of HARQ retransmissions is set to three. 
    For every TTI the effective SINR for each scheduled user is calculated per resource element (RE). Based on those SINR samples per RE for each scheduled UE, the effective SINR is applied for link-to-system-level mapping to determine MMIB \cite{tang2010mean}. The MMIB is calculated as 
    \begin{align}
        \text{MMIB} = \frac{1}{N_{\text{CBS}}} \sum_{b_i\in N_{\text{CBS}}} 1 - H\left( \frac{1}{1+\exp(|\text{LLR}_{b_i}|)}\right),
    \end{align}
    where, $N_{\text{CBS}}$ is the number of bits per each CB, $H(\cdot)$ is the mutual information operator and $\text{LLR}_{b_i}$ is the Log Likelihood Ratio (LLR) of bit $b_i$. Results from link simulations are used to convert the MMIB metric to an error probability to determine if the transmission is successfully decoded or failed \cite{MMIB_metric}. 

\subsection{Performance Results}
    In the following we present results from the dynamic system-level simulations for the following cases:
        \begin{enumerate}
            \item \textbf{\textit{Baseline CQI with TB-based Tx, TBER target=10\%}}: Corresponding to a selection of the highest MCS while not exceeding the 10\% TBER target with TB-based transmission and single-bit HARQ feedback is per TB. Traditional OLLA with 10\% TBER target is assumed.
            \item \textbf{\textit{Baseline CQI with CBG-based Tx, TBER target=10\%}}: Corresponding to the selection of the highest MCS while not exceeding 10\% TBER target for the TB while assuming CBG-based transmissions  and multi-bit HARQ feedback so only failed CBGs are retransmitted. Traditional OLLA with 10\% TBER target is assumed.
            \item \textbf{\textit{eCQI with CBG-based Tx and eOLLA}}: Using the proposed eCQI scheme and the  eOLLA scheme from \cite{GlobeCom_eOLLA} is to control the CBG error probability. 
    \end{enumerate}
    The setting of parameters $N$ and $P$ are selected based on a series of simulations to use the setting that results in the best performance in terms of XR capacity. For the considered XR cases with large TBs typically containing $M=8$ CBGs, we find that a reasonable default setting is $N=4$ and $P=50\%$, meaning that the eCQI corresponds to the maximum supported MCS with at most  out of 8 eight failed CBGs with 50\% probability.

    \begin{figure}[t]
        \centering
            \subfigure[{Experienced CB SINR}]{
                 \centering
                 \includegraphics[width=0.48\textwidth,trim={0 0 0 0},clip] {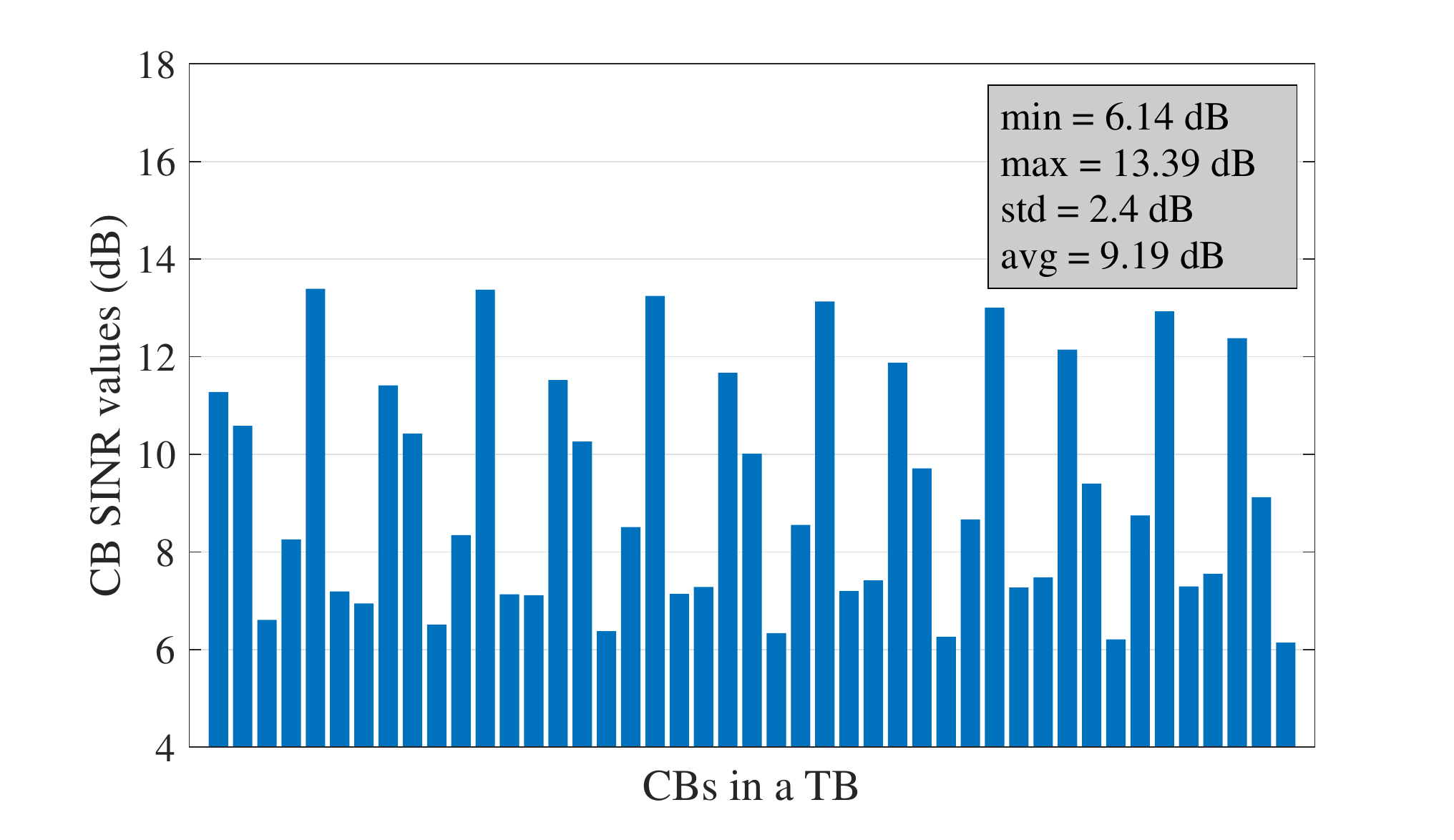}\label{P2:Fig:CB_SINRs}}
             \subfigure[{Experienced CBG SINR}]{
                 \centering
                 \includegraphics[width=0.48\textwidth,trim={0 0 0 0},clip] {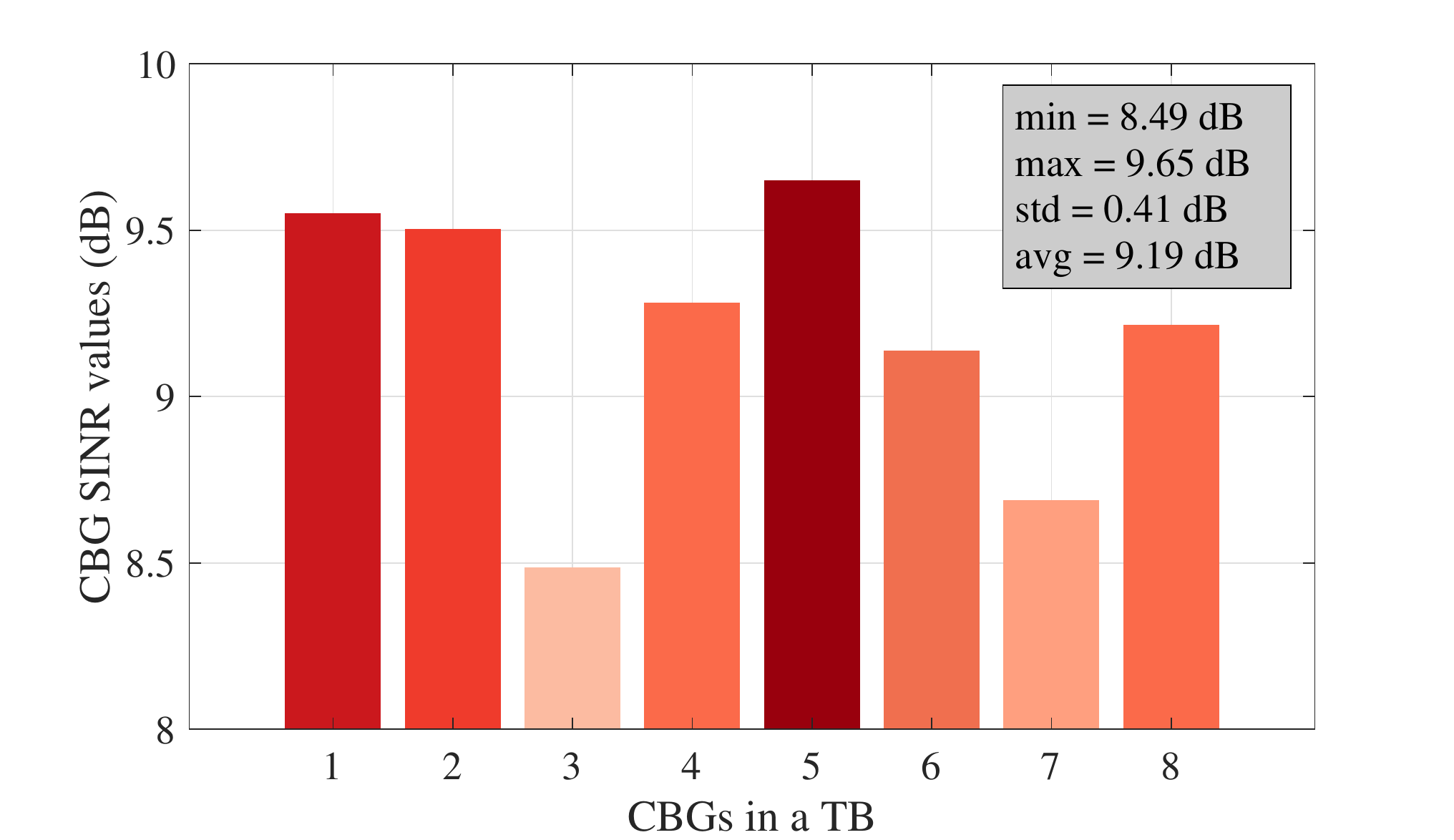}\label{P2:Fig:CBG_SINRs}}  
        \caption{\small{The experienced CB/CBG SINR values for a certain TB of a UE in the network.}}
        \label{P2:Fig:CB_CBG_SINRs}
    \end{figure}
    Fig \ref{P2:Fig:CB_SINRs} shows the experienced CB SINRs (in dB) for the multiple TBs in the first transmissions for a certain UE in the network. As can be seen, there is a correlation among CB SINR values. In addition, the CBs in a TB experience uneven SINR values which result in non-identical CB and CBG error probability. This confirms that we cannot assume i.i.d. errors of the individual CBs (and hence also CBGs), and hence must take correlations and different values of error probabilities for the CBs (and CBGs) into account as expressed in Section \ref{Analytical_Assessment}.

    \begin{figure}[t]
		\centerline{\includegraphics[width=0.65\textwidth,trim={0cm 0cm 0cm 0cm},clip]{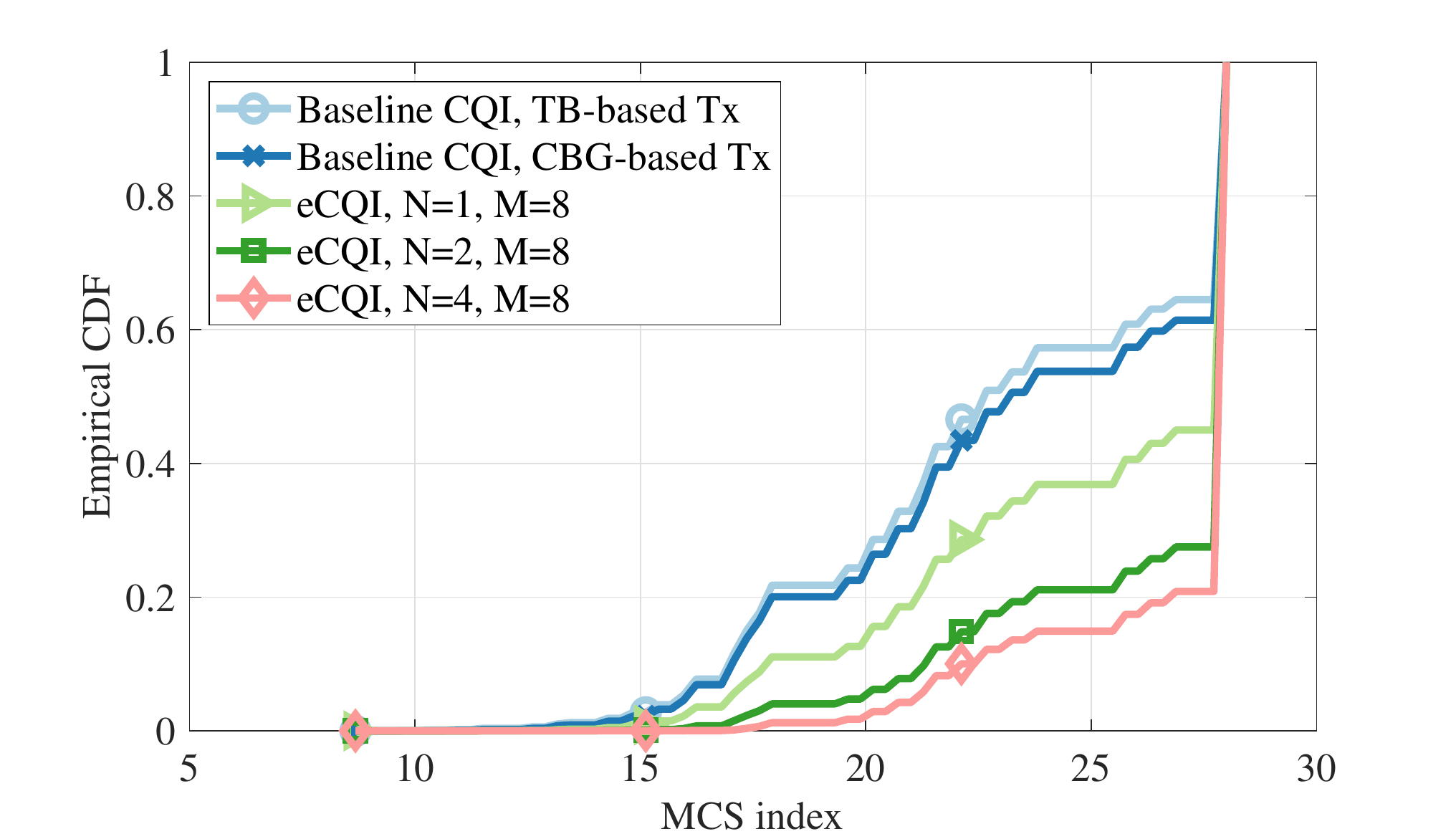}}
		\caption{Empirical CDF of MCS index selection when the minimum required data rate is 45 Mbps and there are 5 UEs per cell.}
		\label{P2:Fig:MCS_index}
    \end{figure}
    Fig. \ref{P2:Fig:MCS_index} shows the empirical cumulative distribution function (CDF) of MCS selection for different schemes, assuming an offered load of 5 XR users per cell with 45 Mbps application layer throughput.  As can be seen, the proposed eCQI method selects higher MCS indices than the baseline cases, resulting in higher radio resource efficiency.  For example, eCQI with $N=4$ selects the highest MCS index 80\% of the time, while the baseline with CBG transmissions chooses the highest MCS index 40\% of the time. This confirms the desired behavior, where gNB selects higher MCS indices for the CBG-based transmissions as this can be afforded, given that still only a fraction of the CBGs will fail in the first transmission, resulting in more resource-efficient transmissions.
	
\begin{figure*}
        \centering
            \subfigure[{Data rate=45 Mbps, 3 UEs per cell}]{
                 \centering
                 \includegraphics[width=0.48\textwidth,trim={0 0 0 0},clip] {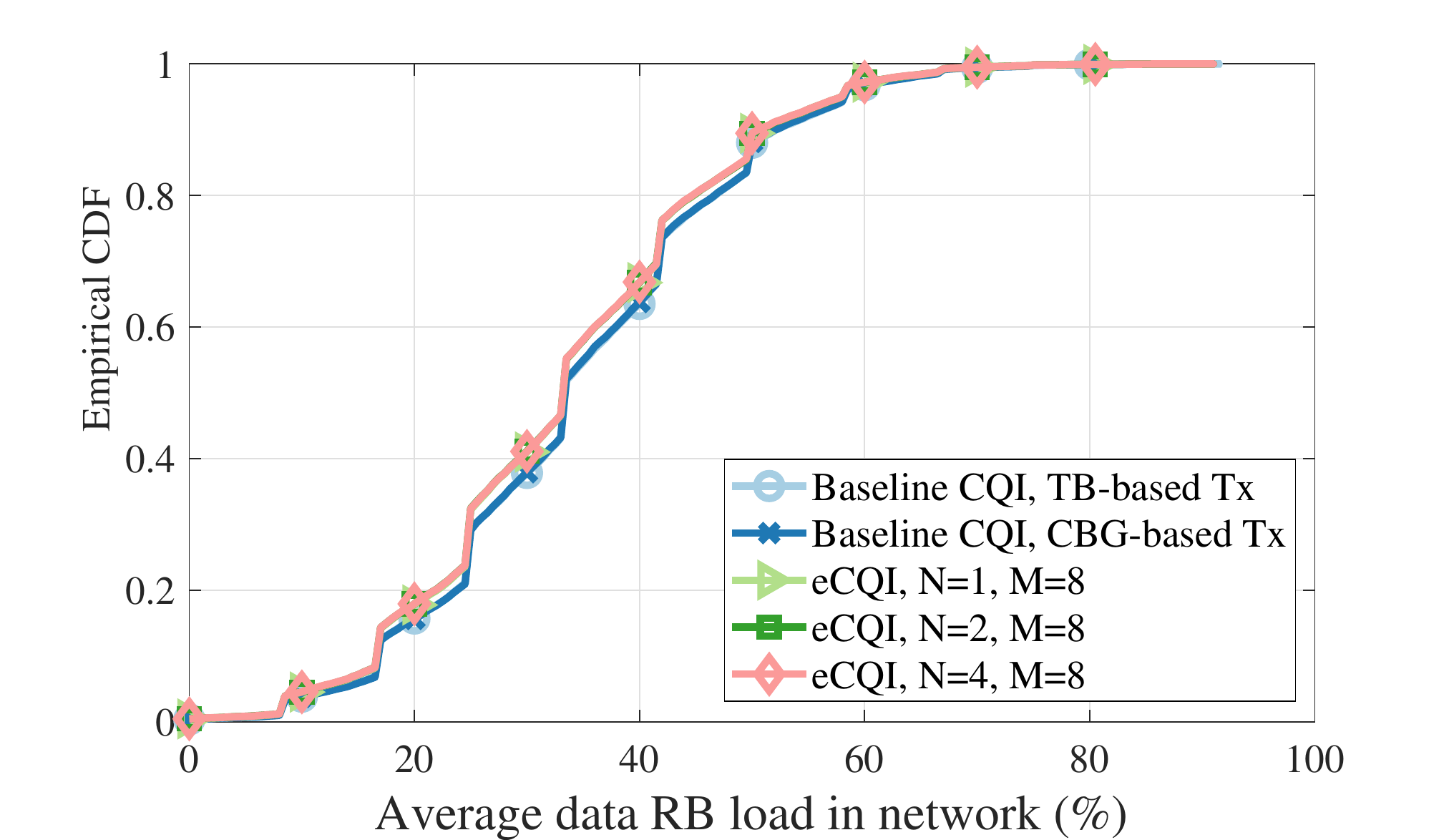}\label{P2:Fig:PRB_Load_45Mbps_3UE}}
             \subfigure[{Data rate=45 Mbps, 5 UEs per cell }]{
                 \centering
                 \includegraphics[width=0.48\textwidth,trim={0 0 0 0},clip] {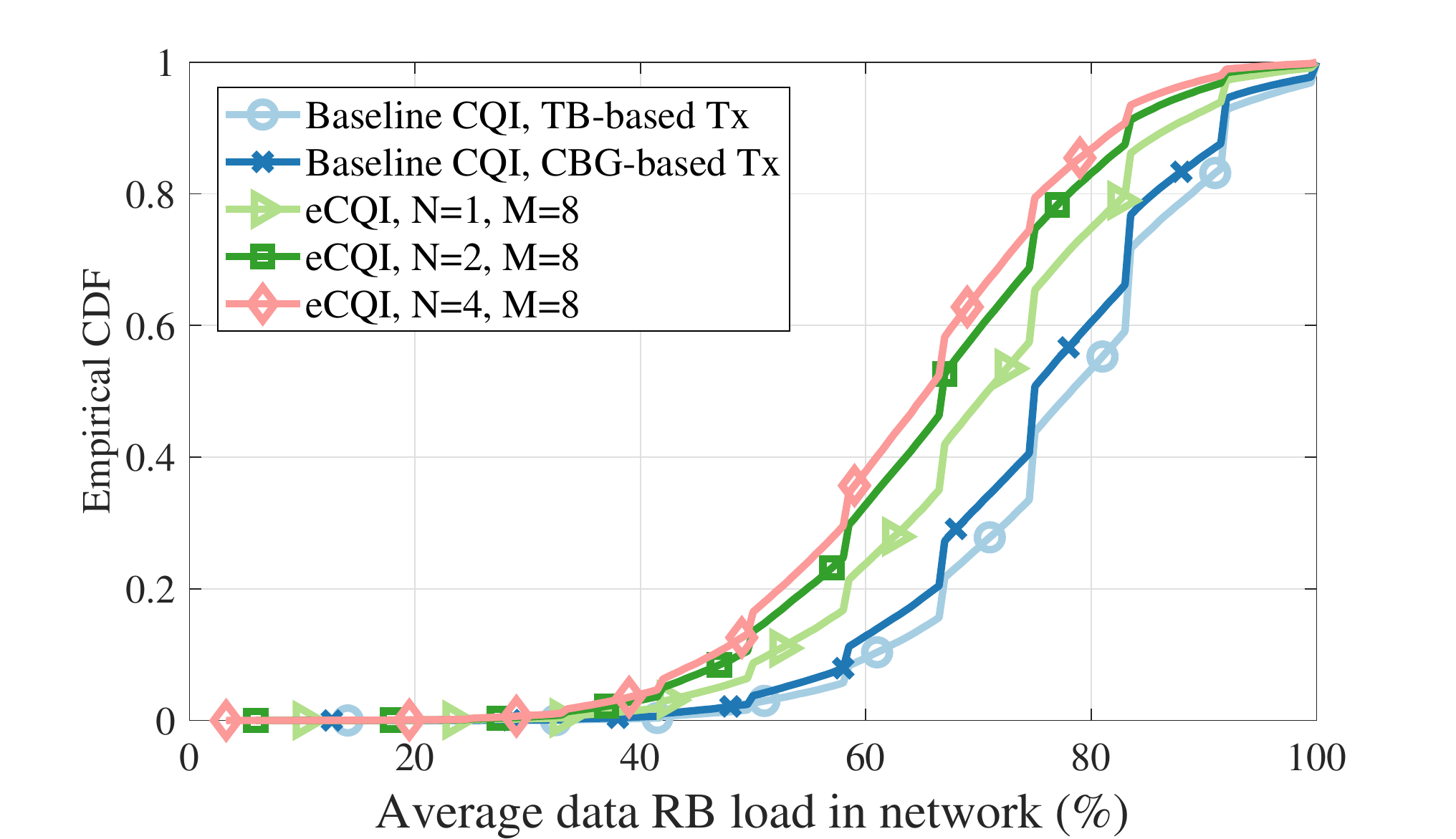}\label{P2:Fig:PRB_Load_45Mbps_5UE}}  
        \caption{\small{Empirical CDF of PRB utilization for the different XR QoS requirements and loads in the network.}}
        \label{P2:Fig:Average_RB_Load}
    \end{figure*}

	Fig. \ref{P2:Fig:Average_RB_Load} shows the empirical CDF of the PRB utilization in the network for offered load levels of 3 and 5 XR users per cell. For the low load scenario, there PRB utilization is displaying only marginal differences, while more significant gains in PRB utilization are observed for the high load scenario, where the eCQI scheme outperforms the baseline cases. The proposed eCQI scheme with $N=4$ improves the resource utilization by 17\% at the 50\textsuperscript{th} percentile as compared to the baseline. This resource efficiency gain originates from partial CBG retransmissions and higher MCS selection as shown in Fig. \ref{P1:Fig:failedCBGs_BLER} and Fig. \ref{P2:Fig:MCS_index}, respectively.
    \begin{figure*}
        \centering
            \subfigure[{Data rate=45 Mbps, 3 UEs per cell }]{
                 \centering
                 \includegraphics[width=0.48\textwidth,trim={0 0 0 0},clip] {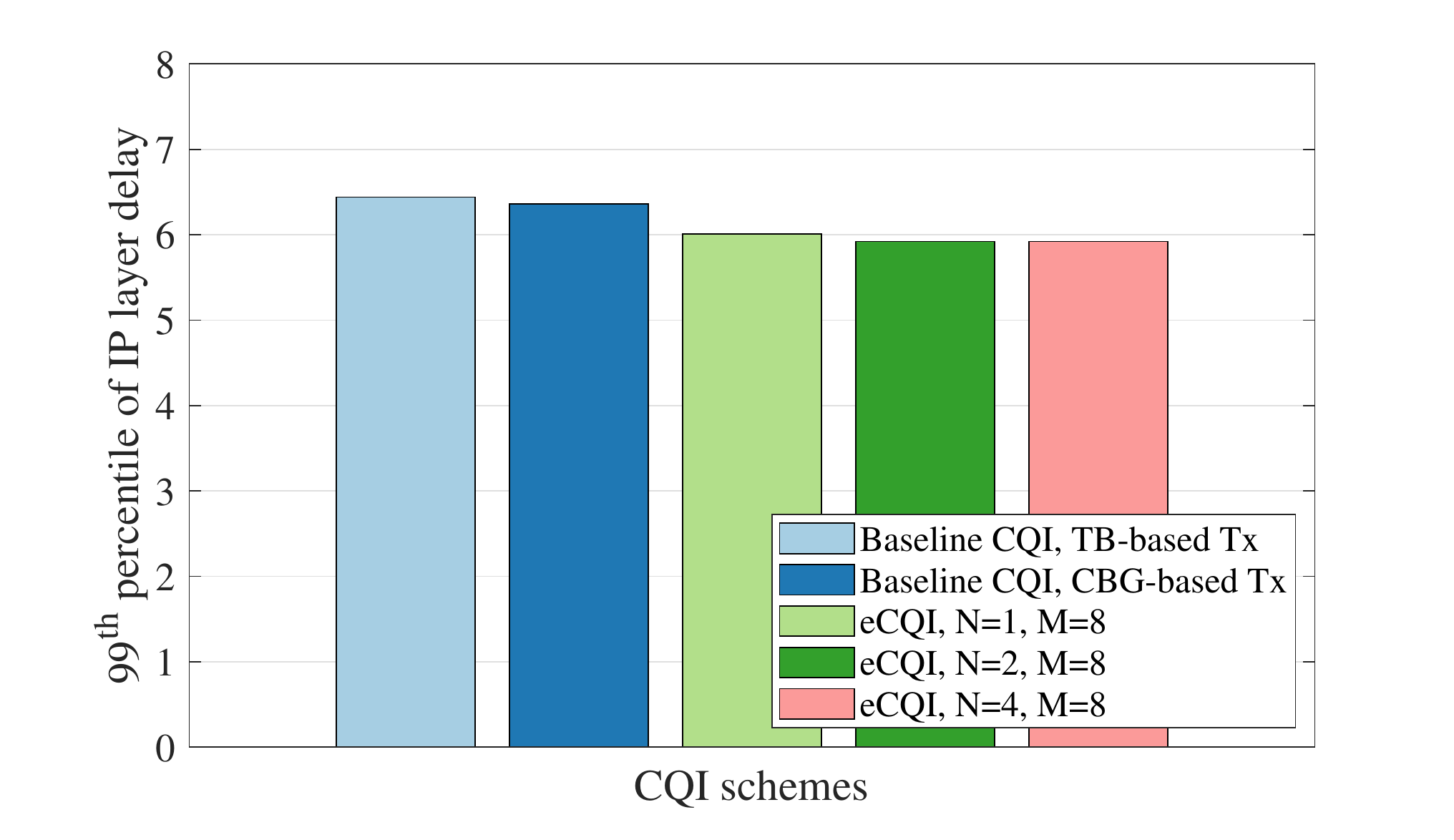}\label{P2:Fig:Delay99_45Mbps_3UEs}}
             \subfigure[{Data rate=45 Mbps, 5 UEs per cell }]{
                 \centering
                 \includegraphics[width=0.48\textwidth,trim={0 0 0 0},clip] {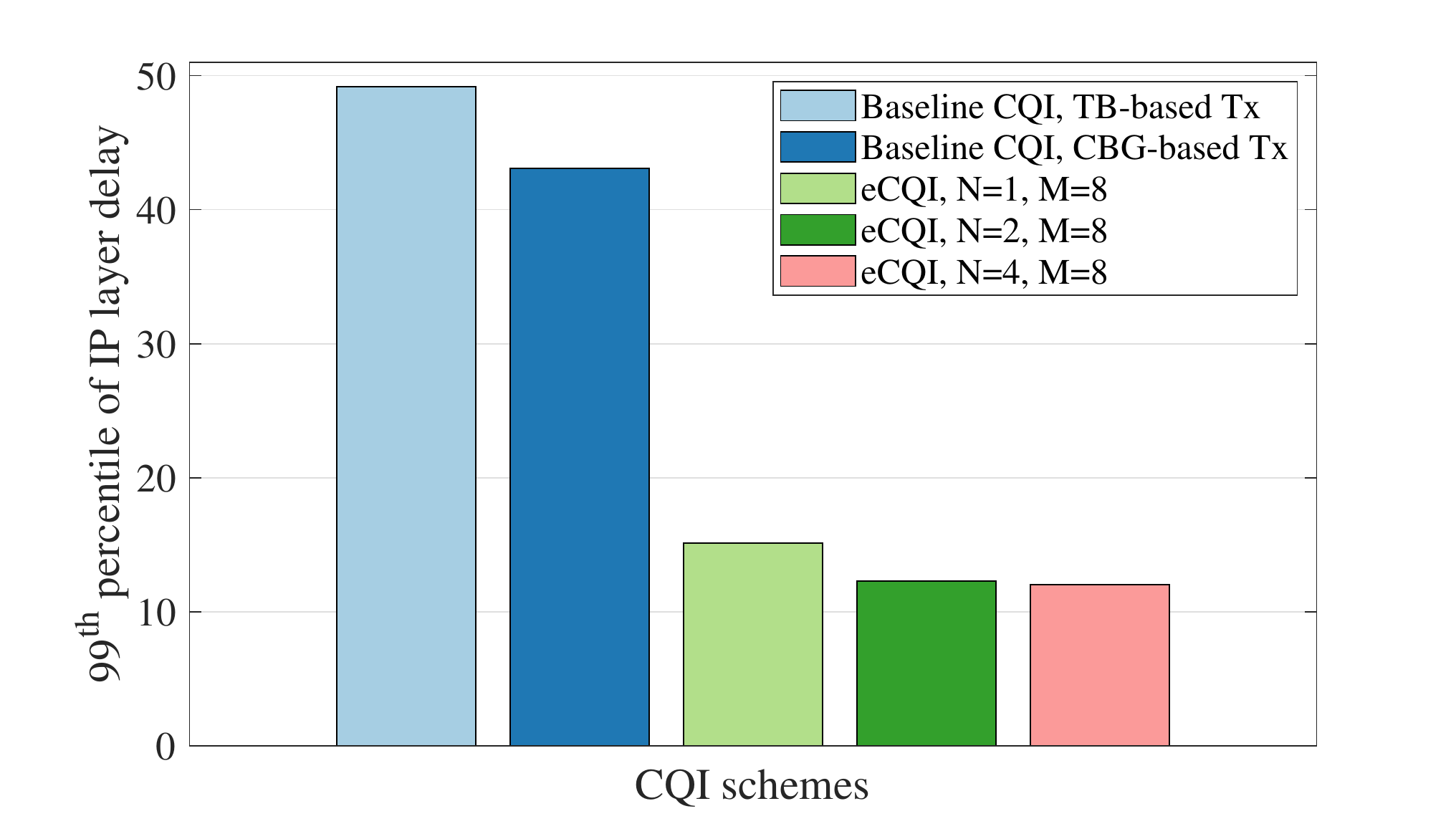}\label{P2:Fig:Delay99_45Mbps_5UEs}} 
                 
        \caption{\small{99\textsuperscript{th} percentile of the empirical CDF of application layer delay for the different XR QoS requirements and loads in the network.}}
        \label{P2:Fig:99percentile_Delay}
    \end{figure*}

       Fig. \ref{P2:Fig:99percentile_Delay} shows the 99\textsuperscript{th} percentile of the empirical CDF of the application layer packet delay for the different load levels with an XR data rate of 45Mbps. It is observed that the delay performance is roughly the same in the low load case, while clear gains are observed for the high load case. This is because temporal variations of queuing at the gNB occur for the high load case especially for the baseline CQI cases, while this happens less often for the eCQI cases as users are served more efficiently on the radio interface.

    \begin{figure*}
        \centering
            \subfigure[{QoS requirement: data rate=30 Mbps, PDB=10ms}]{
                 \centering
                 \includegraphics[width=0.48\textwidth,trim={0cm 0cm 0cm 0.6cm },clip] {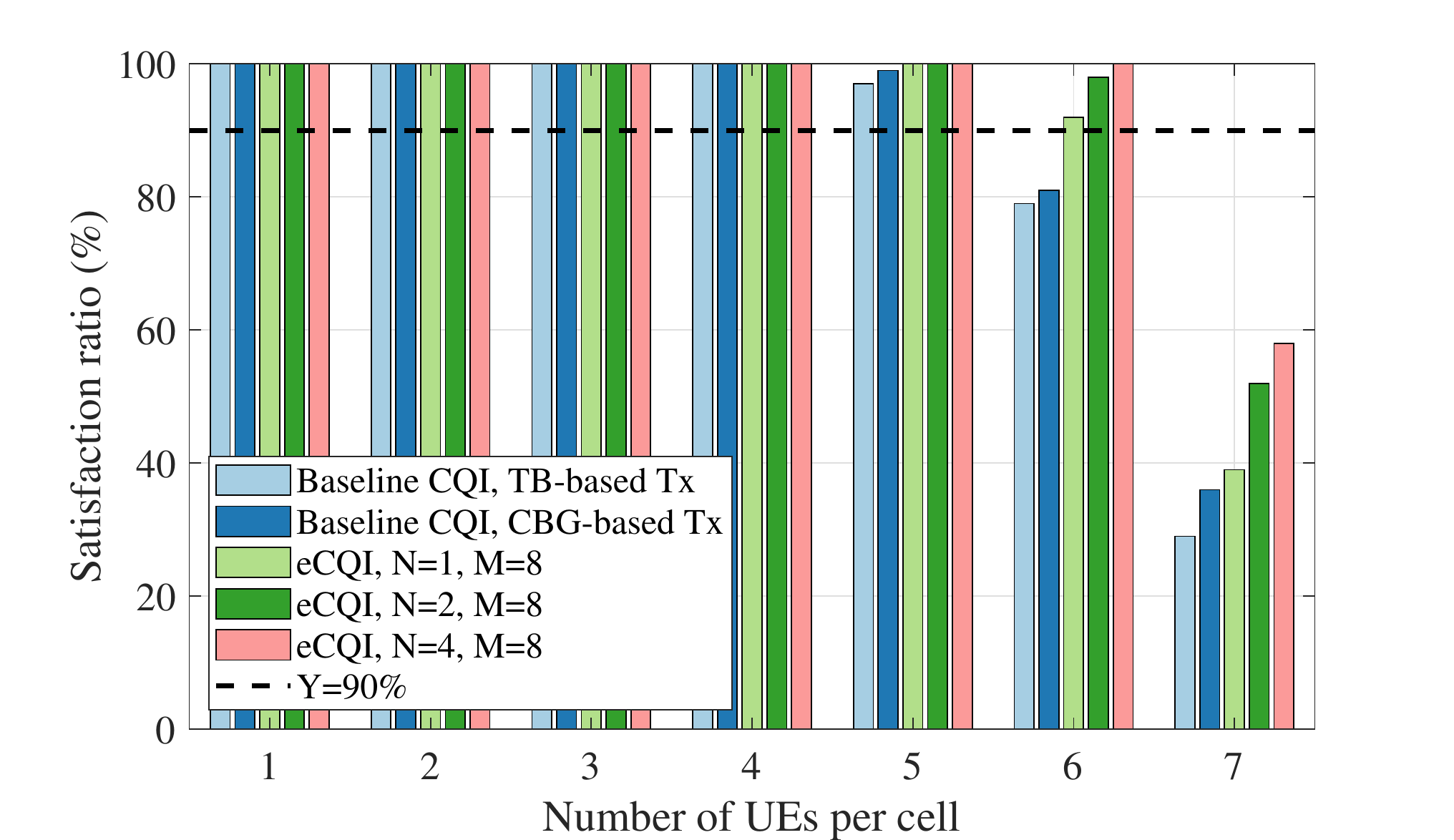}\label{P2:Fig:Happiness_30Mbps_10ms}}
             \subfigure[{QoS requirement: data rate=30 Mbps, PDB=15ms}]{
                 \centering
                 \includegraphics[width=0.48\textwidth,trim={0cm 0cm 0cm 0.6cm },clip] {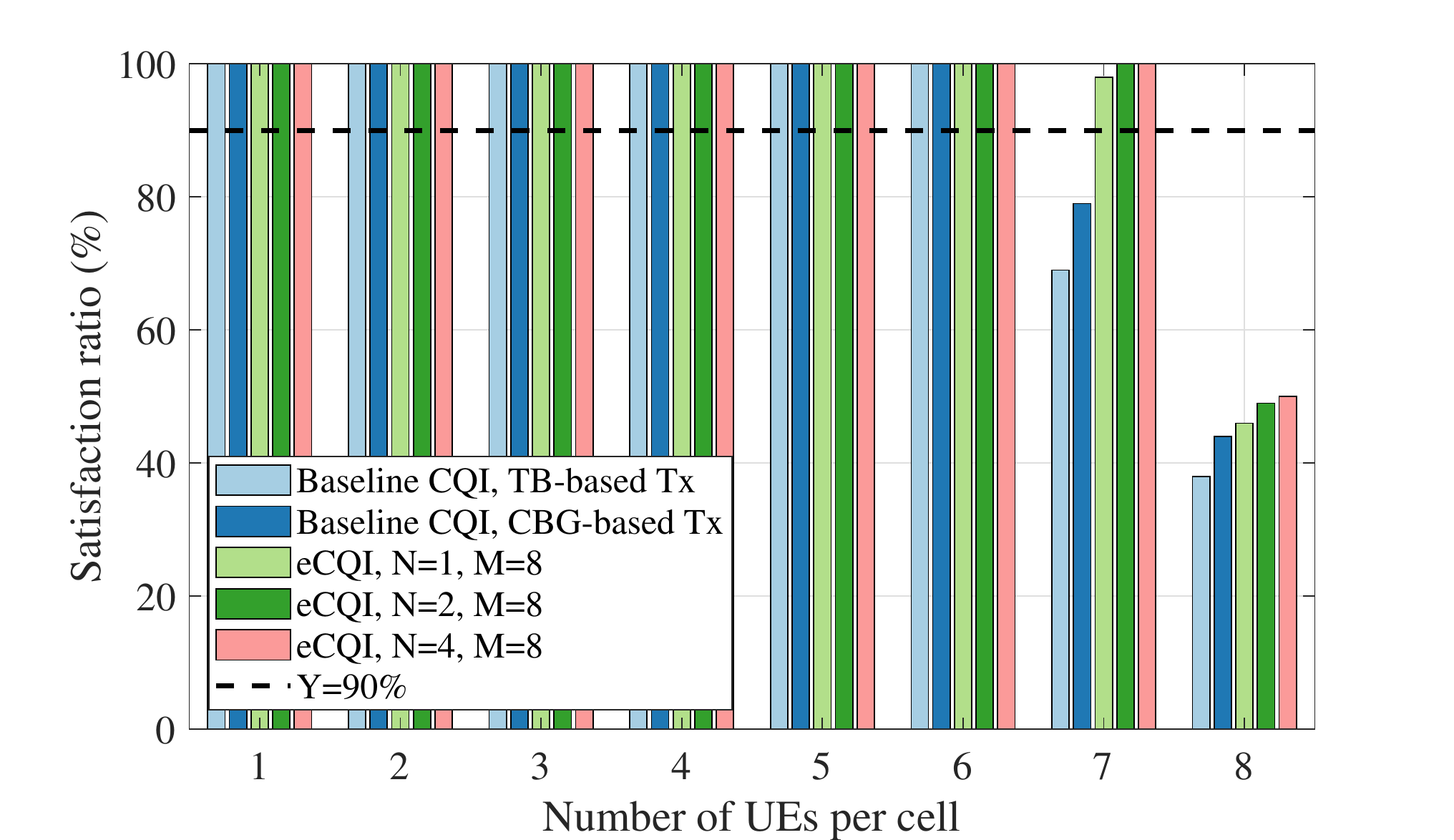}\label{P2:Fig:Happiness_30Mbps_15ms}}  
            \subfigure[{QoS requirement: data rate=45 Mbps, PDB=10ms}]{
                 \centering
                 \includegraphics[width=0.48\textwidth,trim={0cm 0cm 0cm 0.6cm },clip] {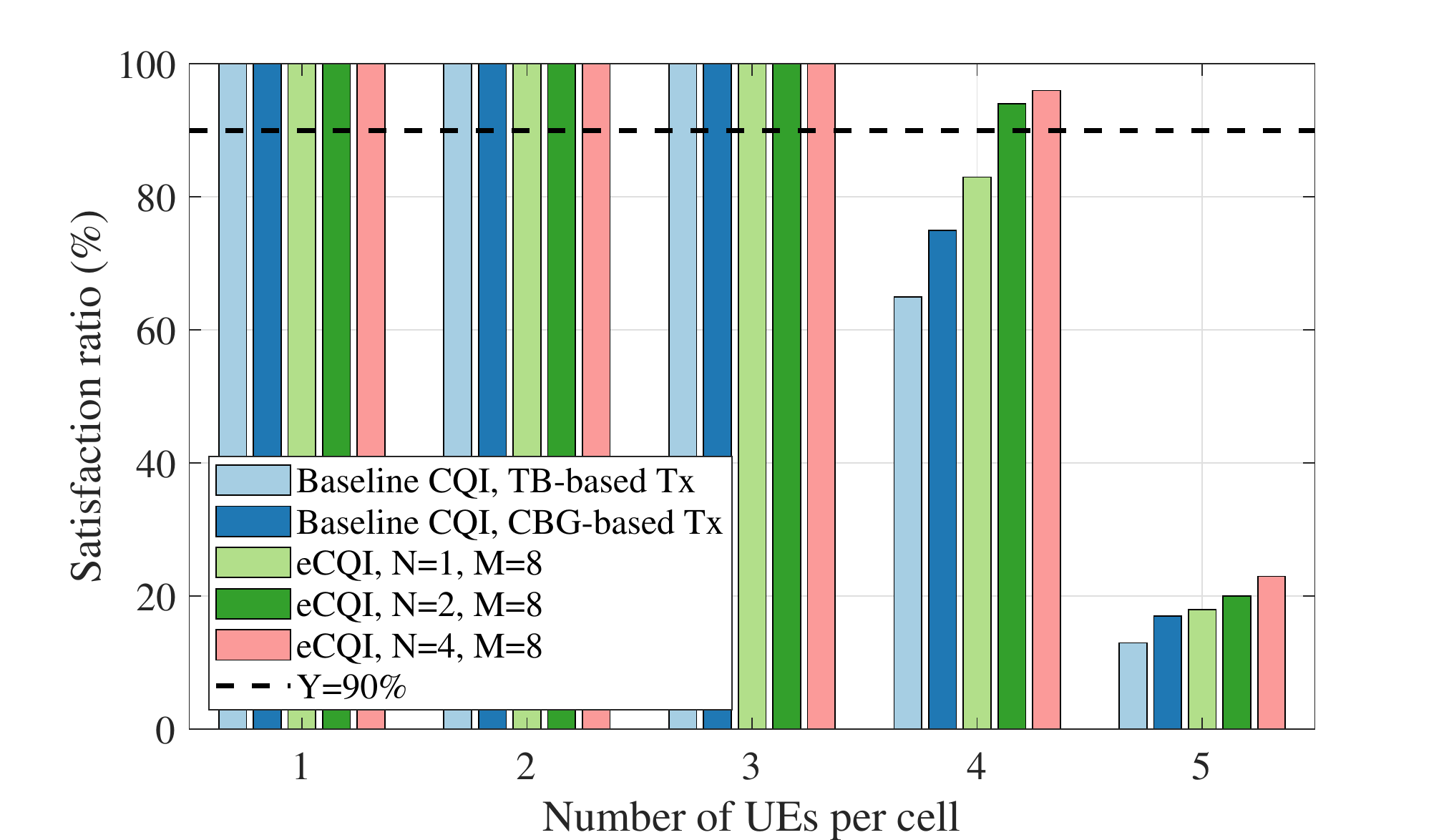}\label{P2:Fig:Happiness_45Mbps_10ms}}
            \subfigure[{QoS requirement: data rate=45 Mbps, PDB=15ms}]{
                 \centering
                 \includegraphics[width=0.48\textwidth,trim={0cm 0cm 0cm 0.6cm },clip] {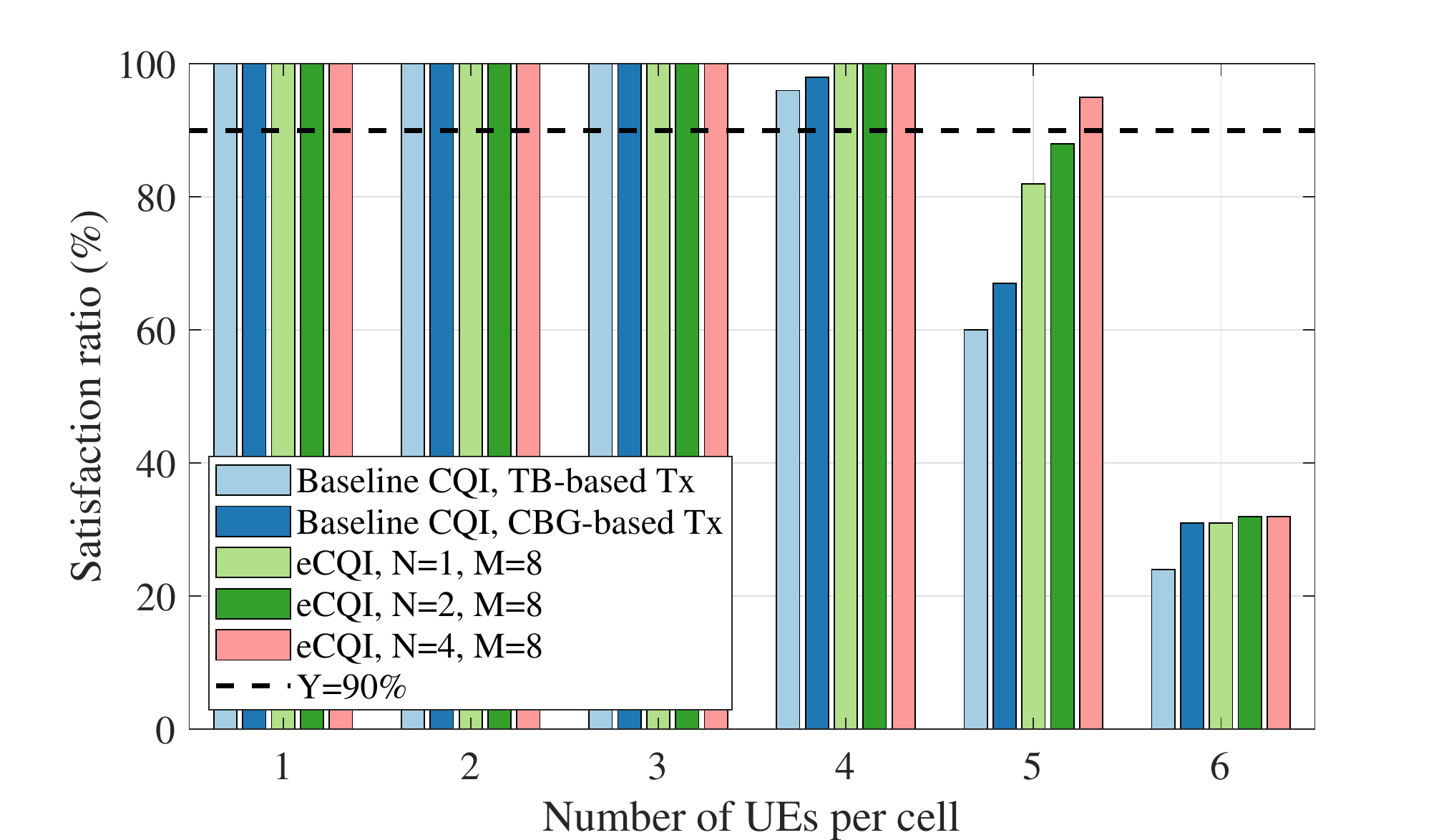}\label{P2:Fig:Happiness_45Mbps_15ms}}     
                 
        \caption{\small{System capacity: the supported number of UEs per cell when at least \SI{90}{\percent} of them are marked as satisfied UEs. A satisfied UE should successfully receive more than \SI{99}{\percent} of its own packets within the PDB \cite{3gpp.38.838}.}}
        \label{P2:Fig:Happiness}
    \end{figure*}

        The main results related to XR capacity assessment are displayed in  Fig. \ref{P2:Fig:Happiness} where the fraction of satisfied users is plotted. The results are shown for different numbers of UEs per cell for four different XR QoS requirements with XR source data rates of 30Mbps and 45Mbps, and PDB values of 10ms and 15 ms. Recall that the XR capacity is defined as the maximum number of supported users per cell, while at least 90\% of those are satisfied. The case 30Mbps corresponds to the so-called "single eye” video streaming, while the 45Mbps case is representing “dual eye” applications. As expected, the higher data rate cases consume more radio resources so less number of UEs can be satisfied per cell. Overall, the proposed eCQI scheme outperforms the baseline cases in all the scenarios. It can be observed that eCQI with $N=4$ supports one more XR UE per cell as compared to the CQI baseline cases. This is equivalent to a system capacity gain of 17\% to 33\% for the different XR scenarios. This is a result of being more resource-efficient when using the eCQI scheme as also pictured in Fig. \ref{P2:Fig:Average_RB_Load}. It is furthermore observed that the XR capacity decreases with few users when the PDB is tightened from 15ms (typical value for cloud gaming applications) to 10ms (typical value for AR and VR cases). The relative XR capacity of using eCQI over baseline CQI is typically higher for the higher XR data rates and lower PDB values, i.e. in the range where there is the most need for improving the otherwise modest XR capacity.
    
\section{Conclusions}\label{Conclusion}
    This paper has introduced an eCQI scheme for 5G-Advanced XR use cases to unleash the full performance potential of CBG-based transmissions. The eCQI provides valuable feedback to the gNB that enables it to conduct more accurate link adaptation, where only up to $N$ out of $M$ CBGs are in error with probability $P$. This complements the recently published eOLLA scheme for also controlling the CBG error rate. 
    
    Our analytical assessment of the CBG-based transmission performance shows significant differences depending on whether individual CBs (and CBGs) have i.i.d. errors, or whether they are correlated and potentially have different error probabilities. Our results from the dynamic system-level simulations show that the i.i.d. assumption is not valid under realistic conditions as we typically observe some correlation and also different error probabilities per CB/CBG as a result of time-frequency variations of the experienced interference per CB/CBG at the receiver. Our XR system-level capacity assessments show attractive gains of 17-33\% from using the eCQI scheme as compared to baseline CQI. The gains are often larger for the more challenging XR use cases with higher application source data rates and tighter PDB. The obtained gains confirm the eCQI benefits of being able to more accurately guide the gNB for proper selection of the MCS, capturing the UE experienced CBG performance characteristics. The complexity at the UE for calculating the eCQI feedback based on radio channel SINR measurements is also assessed. For this purpose, we derived linear closed-form expressions that are easy to implement in the UE for a low complexity eCQI realization. 
    
\appendices
\section{proof of closed-form for \eqref{eq: N eq 2 case}}\label{Appendix A}
\begin{equation}
        \begin{aligned}\label{Proof eq: N eq 2 case}
            &P_e (r,2)  =\!\frac{1}{2!}\!\!\sum_{m_1 \in \mathcal{M}^{(0)}} \!\!\!\!p^r_{m_1}\!\!\!\!\sum_{m_2 \in \mathcal{M}^{(1)}} \!\!\!\! p^r_{m_2}\quad \!\!\!\!\!\!\!\!\bigcap_{m_{3} \in \mathcal{M}^{(2)}}\!\!\Bigl(1-p^r_{m_{3}}\Bigr)
            =\frac{\Pi}{2} \!\!\sum_{m_1 \in \mathcal{M}} O^r_{m_1}\!\!\sum_{m_2 \in \mathcal{M}\setminus m_1} O^r_{m_2} 
            \\&= {\Pi} \!\!\sum_{m_1 \in \mathcal{M}} O^r_{m_1}\Biggl(\sum_{m_2 \in \mathcal{M}\setminus m_1}\!\!\!\! O^r_{m_2} \!+\! O^r_{m_1} \!-\! O^r_{m_1}\!\!\Biggr) 
            = \frac{\Pi}{2} \sum_{m_1 \in \mathcal{M}} O^r_{m_1}\left(\sum_{m_2 \in \mathcal{M}} O^r_{m_2} - O^r_{m_1}\right)
            \\&= \frac{\Pi}{2} \Biggl[\Bigl(\sum_{m_1 \in \mathcal{M}} O^r_{m_1}\Bigr)^2 -  \sum_{m_1 \in \mathcal{M}} \Bigl(O^r_{m_1}\Bigr)^2 \Biggr]
            =\frac{\Pi}{2} \Biggl[\Bigl(M \Bar{O^r}\Bigr)^2 \!- \!M^2 \text{Var}\Bigl(O^r\Bigr) \!+\! \Bigl(M \Bar{O^r}\Bigr)^2 \Biggl]
            \\&= \frac{\Pi M^2}{2}\Bigl(2 \Bar{O^r}^2 - \text{Var}\bigl(O^r\bigr)\Bigr).
        \end{aligned}    
\end{equation}

\section{proof of closed-form for \eqref{eq: N eq 3 case}}\label{Appendix B}
    \begin{equation}
            \begin{aligned}\label{Proof eq: N eq 3 case}
            &P_e (r,3)
            = \frac{\Pi}{3!} \!\!\sum_{m_1 \in \mathcal{M}} \!\!\!\!O^r_{m_1}\sum_{m_2 \in \mathcal{M}\setminus m_1} O^r_{m_2} \!\!\!\!\sum_{m_3 \in \mathcal{M}\setminus m_1,m_2} \!\!\!\!O^r_{m_3}
            \\&= \!\frac{\Pi}{6} \!\!\sum_{m_1 \in \mathcal{M}} \!\!\!O^r_{m_1}
            \!\Biggl[\sum_{m_2 \in \mathcal{M}\setminus m_1} \!\!\!\!\!\!O^r_{m_2}
            \!\Biggl(\sum_{m_3 \in \mathcal{M}\setminus m_1} \!\!\!\!\!\!O^r_{m_3}
            \!\!-\! O^r_{m_2}\!\Biggr)\!\Biggr]
            \\&= \frac{\Pi}{6} \!\!\sum_{m_1 \in \mathcal{M}} \!\!\!O^r_{m_1}
            \!\Biggl[\!\Biggl(\sum_{m_2 \in \mathcal{M}\setminus m_1} \!\!\!\!O^r_{m_2}\!\Biggr)^2 \!\!- \!\!\!\!\!\sum_{m_2 \in \mathcal{M}\setminus m_1} \!\!\Bigl(\!O^r_{m_2}\!\Bigr)^2\!\Biggr]
            \\&= \frac{\Pi}{6} \!\!\sum_{m_1 \in \mathcal{M}} \!\!\!O^r_{m_1}
            \!\Biggl[\!\Biggl(\sum_{m_2 \in \mathcal{M}} \!\!\!O^r_{m_2}
            \!-\! O^r_{m_1}\!\Biggr)^2 \!\!\!
            -\!\!\!\!\!\sum_{m_2 \in \mathcal{M}} \Bigl(O^r_{m_2}\Bigr)^2 
            + \!\bigl(\!O^r_{m_1}\!\bigr)^2\Biggr]\!\!
            \\&= \! \frac{\Pi}{6} \!\!\!\sum_{m_1 \in \mathcal{M}} \!\!\!\! O^r_{m_1}
            \!\Biggl[\!\Bigl(\!\!\sum_{m_2 \in \mathcal{M}} \!\! \!O^r_{m_2}\!\Bigr)^2 
            \!\!-\!\!\!\!\sum_{m_2 \in \mathcal{M}} \!\!\!\!O^r_{m_2}O^r_{m_1}
            + \!\bigl(O^r_{m_1}\bigr)^2 \!- \!\!\!\sum_{m_2 \in \mathcal{M}} \!\!\Bigl(\!O^r_{m_2}\!\Bigr)^2 \!+\! \Bigl(\!O^r_{m_1}\!\Bigr)^2\Biggr]
            \\&=\! \frac{\Pi}{6}\Biggl[\!\Bigl(\!\sum_{m \in \mathcal{M}} \!\!O^r_{m}\!\Bigr)^3 
            - 3 \Bigl(\sum_{m \in \mathcal{M}} O^r_{m}\Bigr)\Biggl(\sum_{m \in \mathcal{M}} \Bigl(O^r_{m}\Bigr)^2\Biggr) + 2 \sum_{m \in \mathcal{M}} \Bigl(O^r_{m}\Bigr)^3\Biggr]
            \\&= \frac{\Pi}{6}\Biggl[\!\Bigl(M\Bar{O^r}\Bigr)^3\!
            \!- 3 M\Bar{O^r}\Bigl(M^2 \Bigl(\Bar{O^r}^2 \!+\!\text{Var}\bigl(\!O^r\!\bigl) \!\Bigr)\!\biggr)
            + 2 \!\!\sum_{m \in \mathcal{M}} \!\!\Bigl(\!O^r_{m}\!\Bigr)\!^3\!\Biggr]
            \\&= \frac{M^3\Pi}{6}\Biggl[-2\Bigl(\Bar{O^r}\Bigr)^3 \!\!- 3 \Bar{O^r}\text{Var}\Bigl(\!O^r\!\Bigr)  \!+\! 2 \!\!\sum_{m \in \mathcal{M}} \Bigl(\!O^r_{m}\!\Bigr)^3\Biggr].
            \end{aligned}
    \end{equation}

\hyphenation{op-tical net-works semi-conduc-tor}
\bibliographystyle{IEEEtran}

\end{document}